\documentclass{aa}
\usepackage{graphics}
\newcommand{\cmtwo}{cm$^{-2}$}  
\newcommand{\cmthree}{cm$^{-3}$}

\newcommand{\pcq}{pc$^{2}$}  
  
\newcommand{\wc}{W cm$^{-2}$}                            
\newcommand{\wcmu}{W cm$^{-2}$ $\mu$m$^{-1}$}

\newcommand{\um}{$\mu$m}                                 
\newcommand{\molh}{H$_{2}$}                              

\newcommand{\water}{H$_{2}$O}

\newcommand{\lsun}{L$_{\odot}$}                          
\newcommand{\msun}{M$_{\odot}$}

\newcommand{\gapprox}{$\stackrel {>}{_{\sim}}$}   
\newcommand{\lapprox}{$\stackrel {<}{_{\sim}}$}
\newcommand{\about}{$\sim$}                       
\newcommand{\powten}[1]{10$^{#1}$}
\newcommand{\av}{$A_{\rm V}$}                     

\newcommand{\roc}{$\rho \, {\rm Oph \, cloud}$}
\newcommand{\scc}{${\rm Serpens \,  cloud \,  core}$}

\newcommand{\nir}{{\sc Nir}}
\newcommand{\fir}{{\sc Fir}}

\newcommand{\esa}{{\sc Esa}} 
\newcommand{\nasa}{{\sc Nasa}} 
\newcommand{\isas}{{\sc Isas}}    

\newcommand{\iras}{{\sc Iras}}

\newcommand{\iso}{{\sc Iso}}

\newcommand{\lws}{{\sc Lws}}
\newcommand{\sws}{{\sc Sws}}
\newcommand{\cam}{{\sc Cam}}
\newcommand{\isocam}{{\sc Isocam}}
\newcommand{\cvf}{{\sc Cvf}}
\newcommand{\fov}{{\sc Fov}}
\newcommand{\hpbw}{{\sc Hpbw}}

\newcommand{\amin}{$^{\prime}$}                   
\newcommand{\asec}{$^{\prime \prime}$}

\newcommand{\radot}[4]{\mbox{#1$^{\rm h}$#2$^{\rm m}$#3$\stackrel{\rm s}
{_{\bf\cdot}}$#4}}  

\newcommand{\decdot}[4]{\mbox{#1$^{\circ}$ #2$^{\prime}$ #3$\stackrel {\prime 
\prime}{_{\bf \cdot}}$#4}}

\newcommand{\asecdot}[2]{\mbox{#1$\stackrel {\prime \prime}{_{\bf \cdot}}$#2}}

\begin{document}

   \thesaurus{08         
              (09.09.1\,\scc;  
               09.03.1;  
               09.07.1;  
               09.04.1;  
               08.06.2)  
               }                
                 
   \title{The ISO-LWS map of the \scc: \\
   		I. The SEDs of the IR/SMM sources\thanks{Based 
  on observations with \iso, an \esa\ project with instruments
  funded by \esa\ Member States (especially the PI countries: France, Germany,
  the Netherlands and the United Kingdom) and with the participation of 
  \isas\ and \nasa.}}


   \author{B. Larsson\inst{1}     	\and
           R. Liseau\inst{1}      	\and
           A.B. Men'shchikov\inst{1} 	\and
	   G. Olofsson\inst{1}    	\and 
	   E. Caux\inst{2}        	\and         
	   C. Ceccarelli\inst{3}  	\and
	   D. Lorenzetti\inst{4}  	\and
	   S. Molinari\inst{5}    	\and 	   
	   B. Nisini\inst{4}      	\and 
	   L. Nordh\inst{1}       	\and 
	   P. Saraceno\inst{6}    	\and
	   F. Sibille\inst{7}     	\and   	   
	   L. Spinoglio\inst{6}   	\and
           G.J. White\inst{1,\,8,\,9}     
	}

   \offprints{B. Larsson}

   \institute{Stockholm Observatory, SE-133 36 Saltsj\"obaden, Sweden
              email: bem@astro.su.se  
  \and  
        CESR CNRS-UPS, BP 4346, F--31028 Toulouse Cedex 04, France
  \and  
	Observatoire Grenoble, 414 rue de la Piscine, BP 53X, F--38041 Grenoble 
	Cedex, France 
  \and
 	Osservatorio Astronomico di Roma, Via Osservatorio 2,
 	I--00040 Monteporzio, Italy 
  \and
  	IPAC/Caltech, MS 100-22, Pasadena, California, USA 
  \and
  	Istituto di Fisica dello Spazio Interplanetario CNR, Tor Vergata, 
  	via Fosso del Cavaliere, I--00133 Roma, Italy 
  \and
  	Observatoire de Lyon, F--69230 St. Genis-Laval, France             
  \and
  	Queen Mary \& Westfield College, Dept. of Physics, University of
  	London, Mile End Road, GB--London E1-4NS, UK
  \and 
  	Astrophysics Group, The Cavendish Laboratory, University of Cambridge, 
  	Madingly Road, Cambridge CB3 OHE, UK 	
   	}

\date{Received date: 6 March 2000\hspace{5cm}Accepted date:}

\titlerunning{\iso-\lws\ map of the \scc}
\authorrunning{B. Larsson et al.}
   \maketitle

   \begin{abstract} \iso-\lws\ mapping observations of the Serpens molecular cloud core are presented.
   The spectral range is $50 - 200$\,\um\ and the map size is 8\amin\,$\times$\,8\amin.
   These observations suffer from severe source confusion at \fir\ wavelengths and we employ
   a {\it Maximum Likelihood Method} for the spectro-spatial deconvolution. The strong and fairly isolated
   source SMM\,1/FIRS\,1 presented a test case, whose modelled spectral energy distribution (SED), 
   within observational errors, is identical to the {\it observed} one. The model results for the other
   infrared and submillimetre sources are therefore likely to represent their correct SEDs. Simulations
   demonstrating the reliability and potential of the developed method support this view.
   
   It is found that some sources do not exhibit significant \fir\ emission and others are most likely not
   pointlike at long wavelengths. In contrast, the SEDs of a number of SMMs are well fit by modified
   single-temperature blackbodies over the entire accessible spectral range. For the majority of
   sources the peak of the SEDs is found within the spectral range of the \lws\ and derived 
   temperatures are generally higher ($\ge 30$\,K) than have been found by earlier deconvolution attempts 
   using \iras\ data. SMM sizes are found to be only a few arcsec in diameter. In addition, the SMMs are 
   generally optically thick even at \lws\ wavelengths, i.e. estimated $\lambda(\tau=1)$ are in the 
   range 160--270\,\um. 
   
   The Rayleigh-Jeans tails are less steep than expected for optically thin dust emission.
   This indicates that the SMMs are optically thick out to longer wavelengths than previously assumed,
   an assertion confirmed by self-consistent radiative transfer calculations. Models were calculated 
   for five sources, for which sufficient data were available, viz. SMM\,1, 2, 3, 4 and 9. These models 
   are optically thick out to millimetre wavelengths (wavelength of unit optical depth 
   900 to 1\,400\,\um). Envelope masses for these SMMs are in the range 2--6\,\msun, which is of course 
   considerably more massive than estimates based on the optically thin assumption.
   The luminosities are in the range 10--70\,\lsun, suggesting the formation of 
   low-mass to intermediate mass stars, so that the existence of such massive envelopes 
   argues for extreme youth of the SMMs in the \scc. 
   
   Finally, we present, for the first time, the full infrared SEDs for the outburst source DEOS, both at
   high and low intensity states.

   \end{abstract}

%

\section{Introduction}

The Serpens molecular cloud is a magnificent laboratory for the study
of multiple low mass star formation. The cloud core has a high density of young
stellar objects (YSOs), including several submillimetre sources: 
for a distance of 260\,pc, Kaas (1999) estimated a surface density of 400 -- 800 YSOs 
per square parsec, which is several times higher than found in the star forming western 
parts of the \roc s (\about\,100\,pc$^{-2}$, Bontemps et al. 2000 and references therein).
The extremely high visual extinction makes the \scc\ nearly exclusively accessible to 
infrared  and sub-/millimetre observations (e.g., Casali et al. 1993, Hurt \& Barsony 1996,
Davis et al. 1999). The submillimetre sources, named SMM by Casali et al. (1993), 
are distributed in a southeast-northwest direction with a concentration of sources in two 
clusters. This is also true for the presumably more evolved objects 
(Class\,I and II, see: Kaas 1999). The region also includes a great number of 
Herbig-Haro objects and molecular flows, where at least some can be connected with SMMs 
(e.g., Ziener \& Eisl\"offel 1999, McMullin et al. 1994, White et al. 1995, Curiel et al. 1996). 
Outflows are seen to accompany star formation at all observed evolutionary stages.

\begin{figure}
  \resizebox{\hsize}{!}{\rotatebox{90}{\includegraphics{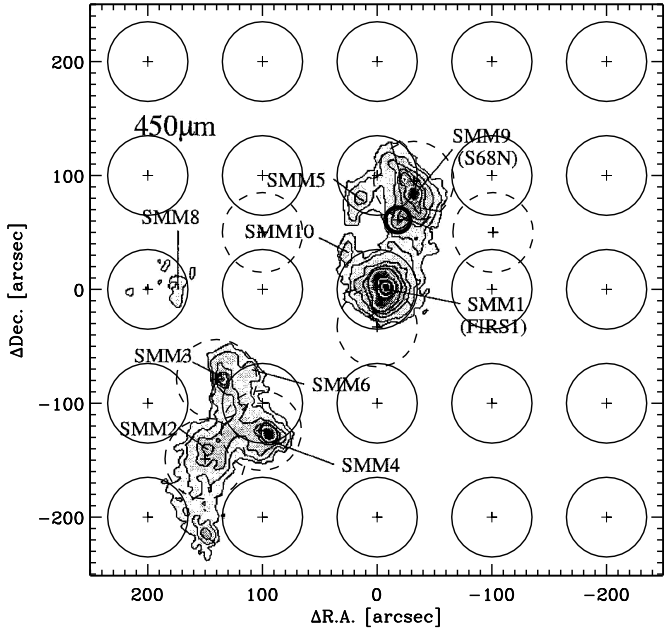}}}
  \caption{Outline of the 8\amin\,$\times$\,8\amin\ map of
  the \lws\ observations, shown together with the 450\,\um\ continuum 
  map of Davis et al. (1999). Conventional source names are indicated.
  Solid circles about grid-points and broken circles about other map-points 
  (see Table~\ref{obstab}) outline the contours of the \hpbw\ of the \lws,
  taken as 70\asec. The small symbol south of SMM9 identifies the pointing
  of the \sws, the aperture of which is rectangular $14^{\prime \prime}\times 
  20^{\prime \prime}$ to $17^{\prime \prime}\times 40^{\prime \prime}$
  depending on the wavelength. The \lws-map is centred on the 
  far-infrared/sub-millimetre source FIRS\,1/SMM\,1 and the offsets of 100\asec\ 
  in both Right Ascension and Declination (plus signs) correspond to the pre-flight 
  estimates of the \lws-beam}
  \label{davispoi}
\end{figure}

\begin{table}
\begin{flushleft}
\caption{\label{obstab} \iso\ observations of the \scc }
  \begin{tabular}{@{}llcrrl@{}}
  \hline
  Target & Principal          & Mode$^{\dagger}$ & 
      \multicolumn{2}{r}{Offsets$^{\ddagger}$ (\asec)} & Date \\
  Name   & Investigator       &      &  R.A.  & Dec.   &      \\
  \hline
  {\bf LWS}    & & & & & \\
  \hline
  Map    & R. Liseau           & L01 &    0   &    0   & 21-Oct-96 \\
  \hline
  Flow   & P. Saraceno         & L01 &    0   &  $-$33 & 23-Oct-97 \\
  Field  & P. Saraceno         & L01 & $-$100 &     50 & 08-Mar-97 \\
  Field  & P. Saraceno         & L01 &    100 &     50 & 08-Mar-97 \\
  SMM2   & M. Spaans           & L02 &    151 & $-$148 & 14-Apr-96 \\
  SMM3   & M. Griffin          & L01 &    140 &  $-$78 & 21-Oct-96 \\
  SMM4   & M. Griffin          & L01 &    101 & $-$124 & 21-Oct-96 \\
  S68N   & M. Spaans           & L02 &  $-$32 &     95 & 14-Apr-96 \\
  \hline
  {\bf SWS}    & & & & & \\
  \hline
  DEOS   & T. Prusti           & S01  & $-$18 &     61 & 14-Apr-96 \\
  DEOS   & T. Prusti           & S01  & $-$18 &     61 & 01-Sep-96 \\
  DEOS   & T. Prusti           & S01  & $-$18 &     61 & 24-Oct-96 \\
  DEOS   & T. Prusti           & S01  & $-$18 &     61 & 08-Mar-97 \\
  DEOS   & T. Prusti           & S01  & $-$18 &     61 & 12-Apr-97 \\
  DEOS   & T. Prusti           & S01  & $-$18 &     61 & 22-Sep-97 \\
  \hline
  {\bf CAM}    & & & & & \\
  \hline
  Serp B & M. Casali           & CVF  &   128 &  $-$54 & 14-Apr-96 \\
  S68N   & P. Andr\'e          & CVF  &    24 &     31 & 14-Apr-96 \\
  DEOS   & G. Olofsson         & LW2  & $-$18 &     61 & 14-Apr-96 \\
  DEOS   & G. Olofsson         & LW3  & $-$18 &     61 & 14-Apr-96 \\
  DEOS   & T. Prusti           & LW2  & $-$18 &     61 & 22-Sep-97 \\
  DEOS   & T. Prusti           & LW3  & $-$18 &     61 & 22-Sep-97 \\
  \hline
  \end{tabular}
\end{flushleft}
Notes to the table: \\
$^{\dagger}$ Instrumental modes: L01 and S01 refer to the full wavelength 
range of the \lws\ and \sws, respectively. L02 stands for \lws\ line spectrum scans. 
CVF denotes spectrum imaging with the Continuous Variable Filter of \isocam.
LW2 and LW3 are two broad band \isocam\ filters centered at 6.7 and 14.3 
\um\ respectively. \\
$^{\ddagger}$ Offsets are relative to the centre coordinates of 
the \lws\ map, viz. R.A.\,=\,\radot{18}{29}{50}{29} and 
Dec.\,=\,\decdot{1}{15}{18}{6}, J\,2000.
\end{table}

Submillimetre sources are often found in dense molecular cloud cores and are believed 
to identify birth places of stars. For the Serpens sources, Hurt \& Barsony (1996) used 
image sharpening techniques to analyse low resolution, broadband \iras\ data.
Cores in low mass star forming regions are generally thought to be cold, with 
temperatures of about 20\,K or below. Emission from such objects will peak at wavelengths 
typically longer than 100\,\um, which makes the Long Wavelength Spectrometer (\lws) 
on board the Infrared Space Observatory (\iso) an instrument more suited than \iras\ 
to study these sources. In addition, the \lws\ provides two orders of magnitude higher 
spectral resolution over its spectral range of 50 to 200\,\um. 

The \lws\ observations presented in this paper aim to address the nature and evolutionary status of 
the SMMs in the \scc, in particular their physical state ($T$, $\rho$, $M$, $L$ etc.). 
The observables to be discussed are the spectral energy distributions of these objects,
obtained over a broad spectral range. This paper utilises mapping observations 
of the \scc\ with the \lws\ and complementary observations, including pointed \lws\ data and 
observations with other instruments aboard \iso. In Sect.\,2, an account of these observations
(8\amin\,$\times$\,8\amin\ map and \iso\ archive retrievals) and their reduction 
is given, and the results are presented in Sect.\,3. The \iso-\lws\ map is highly undersampled and 
in Sect.\,4, we introduce a {\it Maximum Likelihood Method} for the extraction of the SEDs 
of the spatially confused submm sources. The discussion focusses on the comparison
of conventionally applied analysis methods (single temperature, optically thin approximation) 
with that based on the detailed and self-consistent modelling of the radative transfer. 
Finally, in Sect.\,5, the main conclusions of this work are briefly summarised.

\section{Observations and data reductions}

\subsection{The ISO observations}

A grid of 25 positions of \fir-spectra in the \scc\ was obtained with the 
Long-Wavelength Spectrometer (\lws; 43 -- 197\,\um, $R_{\lambda}=140 - 330$) 
on board the Infrared Space Observatory (\iso) on October 21, 1996. 
An account of the \iso-project is given by Kessler et al. (1996).
The \lws\ is described by Clegg et al. (1996) and Swinyard et al. (1996).

The formal map centre, viz. $\alpha$ = \radot{18}{29}{50}{29} and $\delta$ = \decdot{1}{15}{18}{6}, 
epoch J\,2000, coincides to within 10\asec\ with the position of the 
sub-millimetre source SMM\,1. The pointing accuracy in the map is determined as 1\asec\ (rms). 
The spacings between positions in the map, oriented along the coordinate axes, are 
100\asec\ in both Right Ascension and Declination. These offsets correspond 
to the pre-flight estimates of the circular \lws-beam, \hpbw. The actual beam widths
are however smaller, 70\asec\ to 80\asec, leading to appreciable spatial undersampling. 
This is also evident in Fig.\,\ref{davispoi}, where the \lws-beams and pointings are 
represented by circles. The size of the $5\times 5$ \lws-map is thus
8\amin\,$\times$\,8\amin, corresponding to 
$(0.7\times 0.7 = 0.5)$\,$D_{310}^2$\,\pcq, where $D_{310}$ denotes the distance
to the \scc\ in units of the adopted value of 310\,pc (de\,Lara et al. 1991). 

At each map-point the grating of the \lws\ was scanned 6 times in fast mode, oversampling 
the spectral resolution by a factor of 4. Each position was observed for nearly 15\,min.
The centre position was re-observed half a year later on April 15, 1997, for a considerably 
longer integration time (24 spectral scans; Larsson et al. in prep.). 
The internal agreement between the deep integration and the map 
spectrum is excellent, giving confidence that the rest of the map data is also
of good quality.

\begin{figure}
  \resizebox{\hsize}{!}{\rotatebox{90}{\includegraphics{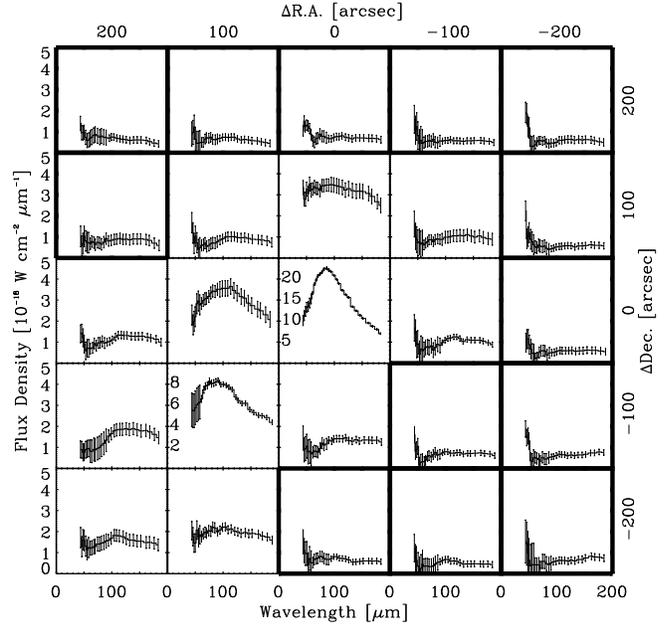}}}
  \caption{50 -- 200\,\um\ spectra towards the grid points of the \lws-map.
  The spatial offsets are given along the upper and right-hand axes.
  The wavelength scale is found along the lower axes and that of the flux density 
  along the left-hand axes, except for (0\asec, 0\asec) and (--100\asec, --100\asec),
  for which the scale is given inside the sub-frames. The \lws\ data have
  been rebinned to the spectral resolution $R_{\lambda}=20$. The error bars
  reflect the differences of the displayed median data (solid line) and the 
  original data, prior to the detector stitching. The \fir-background spectrum is 
  defined as the straight average of the spectra inside the boxes with the thick lines}
  \label{rawdata}
\end{figure}

Our map spectra were complemented with \lws\ spectra at 7 positions inside
the mapped area, which were retrieved from the \iso-archive (Table\,\ref{obstab}). 
In addition, 1D-spectral observations with the Short-Wavelength Spectrometer 
(de\,Graauw et al. 1996: \sws; \fov\,=\,$14^{\prime \prime}\times 20^{\prime \prime}$ to 
$17^{\prime \prime}\times 40^{\prime \prime}$, 2.4 -- 45\,\um,  
$R_{\lambda} \sim 10^2 - 10^3$) and imaging spectrophotometry with the Continuous Variable Filter 
(\cvf; 2.5 -- 15.5\,\um, pixel-\fov\,=\,6\asec, $R_{\lambda}$\,\gapprox\,35) of \isocam\ 
(Cesarsky et al. 1996) were also analysed. Further, broadband \isocam\ images at 6.7
and 14.3\,\um\ and with pixel-\fov\ both \asecdot{1}{5} and 3\asec\ were part of the observational 
material.

In summary, Table~\ref{obstab} provides an overview of the observational material 
employed in this paper. Fig.\,\ref{davispoi} shows the \sws\ and \lws\ pointing positions 
superposed onto the sub-millimetre continuum map (450\,\um) of Davis et al. (1999),
where 9 discrete sub-millimetre sources (SMM\,1 -- SMM\,10, except SMM\,7) are identified.

\begin{figure}
  \resizebox{\hsize}{!}{
  \rotatebox{90}{\includegraphics{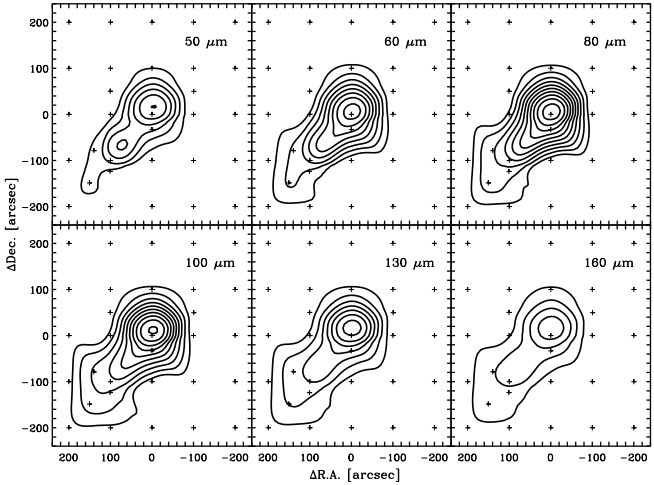}}
                        }
  \caption{Spatial distribution of the continuum emission in the \lws-map 
  in wavebands centred on 50, 60, 80, 100, 130 and 160\,\um, with spectral
  bandwidth of 10\,\um. Contours are spaced by 
  2\,\powten{-17}\,\wc, the lowest contour is at 2\,\powten{-17}\,\wc\
  and the displayed dynamic range is a factor of 10}
  \label{lwscontmap}
\end{figure}

\begin{figure}
  \resizebox{\hsize}{!}{
  \rotatebox{90}{\includegraphics{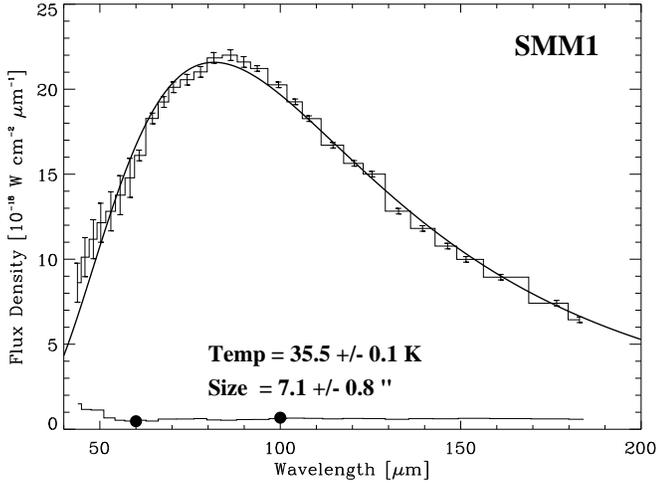}}
                        }
  \caption{A pure blackbody fit ($\beta = 0$, see the text) to \lws\ data of SMM\,1 
  results in the indicated values of the temperature and diameter of a circular source,
  the formal errors of which are given as the standard deviation about the best-fit values 
  ($\chi^2$-minimisation). The displayed \lws\ spectrum has been re-binned to a resolution of 20. 
  The displayed error bars correspond to the relative flux uncertainty, 
  when stitching together the 10 individual detectors of the \lws. At the bottom of the figure,
  the (subtracted) average background spectrum is shown for reference and compared to \iras\ data 
  (filled circles)}
  \label{smm1bbfit}
\end{figure}

\subsection{The reduction of the ISO data}

The \lws\ observations were pipeline-processed (OLP\,7.1) and were subsequently
reduced with the interactive analysis package LIA. Post-pipeline processing 
of the \sws\ and \lws\ data was done with the package ISAP and of the
\isocam\ data with the corresponding CIA programs.  

At each position, the individual \lws\ scans for each of the 10 detectors were 
examined, `deglitched' and averaged. Corrections were applied to the
`fringed' spectra in the map. The fringing indicates that the emission is
extended and/or that point sources were not on the optical axis of the \lws, i.e.
the radial distance from the optical axis was larger than about 25\asec.

For the \lws\ data we believe that the absolute flux calibration is good
to an accuracy of 30\% (Swinyard et al. 1996), whereas relative offsets between
overlapping spectral regions of adjacent detectors were generally within 10\%
(`detector stitching' uncertainty, see also Fig.\,\ref{rawdata}). Internally, the 
\lws-accuracy is much higher than 30\% for the \scc, as is evidenced by observations 
of the same targets at different times and under different observing conditions 
(e.g., pointed vs. mapping observations). 

The absolute accuracy could be less, however, in the very weak spectra towards the 
northern and western edges of the map, for which the
flux uncertainties are dominated by the dark current correction. This was estimated
based on measurements directly before and after, but not during, the mapping 
observations. Suspected dark current drifts during the observation of the first 5 map points, 
viz. from (--200\asec, --200\asec) to (--200\asec, 200\asec), were however approximatively 
linear in time, whereas for the remaining spectra, the dark current seemed to
have stabilised at a constant value. Support for our applied dark current corrections is
provided by the fact that these resulted in fluxes which are in very good agreement with 
IRAS-ISSA data at 60 and 100\,\um\ (formally within 7\% and 4\%, respectively, see: 
Fig.\,\ref{smm1bbfit}).

The reduction of the \sws\ data is similar to that of the \lws\ data, whereas the 
reduction of the \cvf\ data involved additional corrections for transients and flat field.

\section{Results}

From Fig.\,\ref{rawdata} it can be seen that detectable \fir\ emission is present 
at all observed positions. The sensitivity of these map data
at long \lws\ wavelenghts is of the order of \powten{-19}\,\wcmu\ (about 3\,Jy per
beam at 100\,\um), whereas this noise level becomes worse by a factor of about two 
towards the shorter wavelengths.

As pointed out in Sect.\,2.2, most of the observed spectra were fringed.
The situation is different at the map centre, where the \lws\ spectrum shows hardly 
any fringes at all. This strongly suggests that the \fir-emission is dominated by a 
source which is pointlike to the \lws\ and which was reasonably well centred in the
aperture during the observation. As is also evident from Fig.\,\ref{lwscontmap}, 
where \lws\ measurements in 6 continuum bands between 45\,\um\ and 165\,\um\ 
are presented, that this source dominates the emission at all \fir\ wavelengths.
In the following section, the physical characteristics of this source will be 
discussed.

\section{Discussion}

\subsection{The sub-millimetre source SMM\,1}

\subsubsection{The blackbody observed by the LWS}

The centre position of the map dominates the emission over the entire wavelength 
range of the \lws\ (Fig.\,\ref{lwscontmap}). This map point coincides 
with the strong sub-millimetre source SMM\,1/FIRS\,1. At \fir\ and sub-mm wavelengths,
this source is relativly isolated. Any problem of source confusion is therefore 
very much reduced in this case. 

\begin{figure*}
  \resizebox{\hsize}{!}{
  \rotatebox{90}{\includegraphics{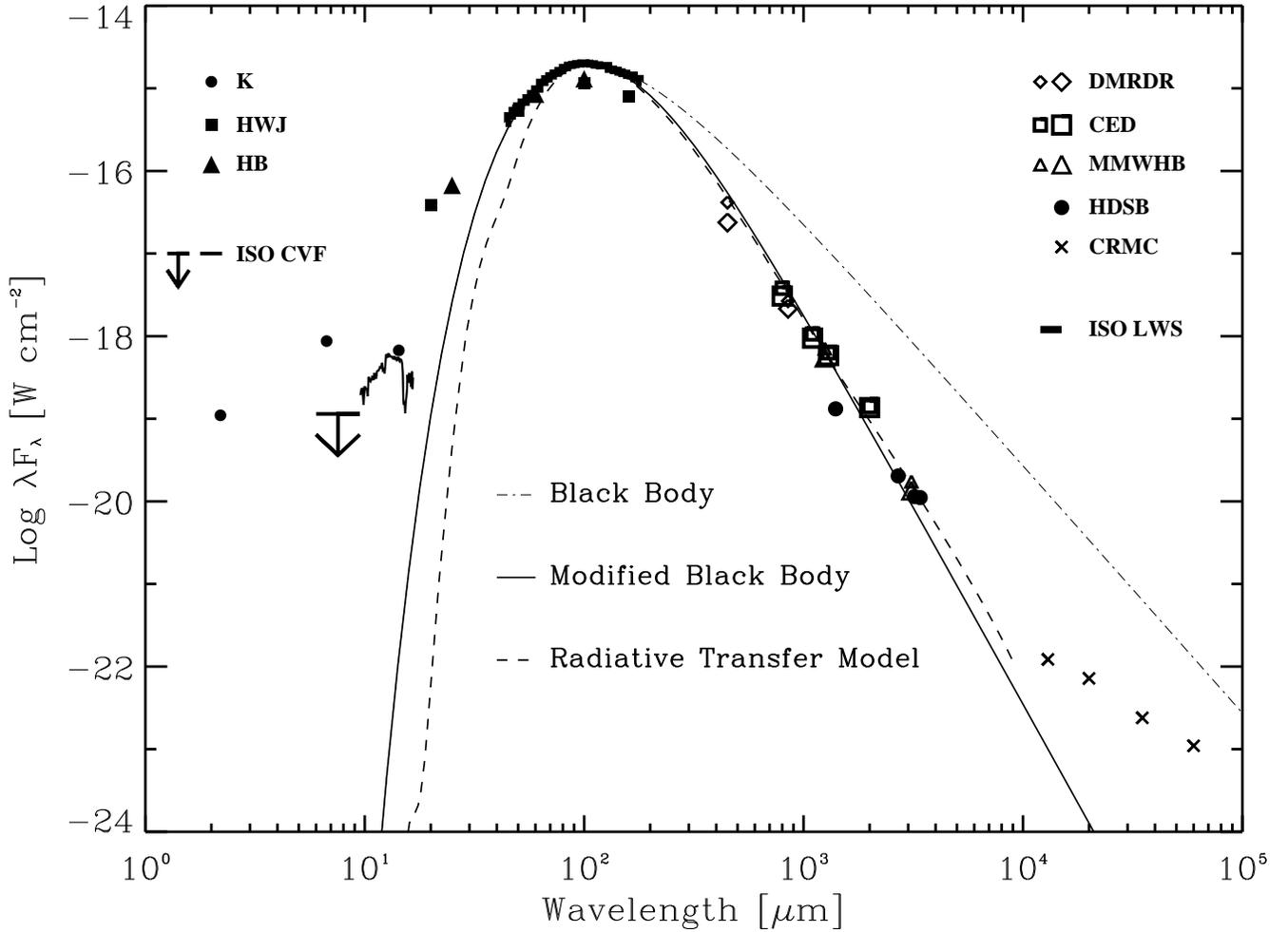}}
                        }
  \caption{The SED ($\lambda\,F_{\lambda}$ vs $\lambda$) of SMM\,1, between 
  2\,\um\ and 6\,cm. Open symbols identify data which have been obtained 
  with apertures larger than 7\asec, with the small open symbols indicating the 
  according to Eq.\,(3) beam-matched data. The thin solid line shows a conventional (modified 
  blackbody) fit to these data within 40\,\um\,\lapprox\,$\lambda$\,\lapprox\,3\,mm and,
  for reference, the 36\,K-blackbody of Fig.\,\ref{smm1bbfit} is shown by the dash-dot line. 
  The dashed line depicts the SED obtained from a model, for which the radiative transfer 
  through the dust envelope is treated self-consistently.
  The line thickness for the \lws\ and \cvf\ data, respectively, are shown in the 
  legends inside the figure. The accompanying upper limit symbol refers to the
  short-wave \cvf\ data. The photometric data collected from the literature are
  referenced to as K = Kaas (1999), HWJ = Harvey et al. (1984), HB = Hurt
  \& Barsony (1996), DMRDR = Davis et al. (1999), CED = Casali et al. (1993),
  MMWHB = McMullin et al. (1994), HDSB = Hogerheijde et al. (1999), CRMC = Curiel et al.
  (1993). The near-infrared data ($\lambda$\,\lapprox\,10\,\um) are probably 
  unrelated to SMM\,1 (see: the text)}
  \label{smm1modbbfit}
\end{figure*}

As evident in Fig.\,\ref{smm1bbfit}, the \fir\ spectrum is well fit by a 
single blackbody of temperature 36\,K. Such a relatively high temperature was also
determined by, e.g., Nordh et al. (1982), Harvey et al. (1984), McMullin et al. (1994) 
and Davis et al. (1999). The diameter of the hypothetically circular source is 
\about\,7\asec, i.e. $(2\,200 \pm 250)\,D_{310}$\,AU.  
This latter value confirms the point-like nature of the far-infrared source, as was
suspected in the previous section on the basis of its largely unfringed spectrum. 
It is also comparable to the source sizes seen in the interferometric observations 
at millimetre wavelengths by Hogerheijde et al. (1999). In contrast, the temperature derived 
here is significantly higher than the 27\,K obtained by Hurt \& Barsony (1996) from high resolution 
\iras\ data (HIRES). 

\subsubsection{A modified blackbody fit to the SED of SMM\,1}

In Fig.\,\ref{smm1modbbfit}, the \lws\ spectrum is shown together with data collected
from the literature. From this overall Spectral Energy Distribution (SED) it is 
clear that the \lws\ data are sufficient to determine the luminosity of the source. 
However, it is also apparent that at longer wavelengths the SED is steeper than the 
Rayleigh-Jeans fall-off of the blackbody ($\lambda\,F_{\lambda} \propto \lambda^{-3}$), which
could indicate that the emission is becoming optically thin somewhere outside the 
spectral range of the \lws.
Provided that in this optically thin regime the dust opacity follows a power law, viz. 
$\kappa_{\lambda} \propto \lambda^{-\beta}$, this index $\beta$ can be determined 
from the observations, since then $\lambda\,F_{\lambda} \propto \lambda^{-(3 + \beta)}$
as $\lambda \rightarrow \infty$
for any density and/or temperature structure of the envelope (spherical symmetry and
central radiating source). A least squares fit to the data between 0.8\,mm and 3.4\,mm yields 
$\beta_{\rm obs} = - \left ( \frac {d \log \lambda F_{\lambda}} {d \log \lambda} + 3 \right ) 
= 1.2 \pm 0.2$. At these relatively long wavelengths, this statistical error is considerably 
larger than any deviation of the modified Rayleigh-Jeans law from the (modified) Planck curve, viz. 
$F_{\lambda}({\rm Planck})/F_{\lambda}({\rm RJ}) = 
\left (1.44\times 10^4/\lambda_{\mu {\rm m}}\,T \right )/\left [ \exp\,(1.44\times 10^4/\lambda_{\mu {\rm m}}\,T) 
- 1 \right ]$, which for $T=36$\,K would result in the ``correction'' of $\beta_{\rm obs}$ by $-0.087$.

Following standard procedures (e.g. Hildebrand 1983, Emerson 1988), we can then
estimate the envelope mass. Using only the data between 
40\,\um\ and 3\,mm, which were collected with sufficiently large telescope beams, 
i.e. larger than 7\asec\ and shown as open symbols in Fig.\,\ref{smm1modbbfit}, 
a modified blackbody can be fit to this section of the SED for a single temperature. 
Hence, the flux density is given by

\begin{equation}
F_{\lambda} = \left(1\,-\,e^{-\tau_{\lambda}}\right)\,\Omega_{\lambda,\,S}\,
B_{\lambda}\left(T\right)
\end{equation}

with obvious notations. In this wavelength regime, the opacities of dust grains 
generally display power law dependencies (e.g., Ossenkopf 
\& Henning 1994). Hence, let the continuum optical depth scale as  

\begin{equation}
\tau_{\lambda} = \tau_{\lambda,\,0} \left(\frac{\lambda_0}{\lambda}\right)^{\beta}
\end{equation}

\begin{figure*}
  \resizebox{\hsize}{!}{
  \rotatebox{90}{\includegraphics{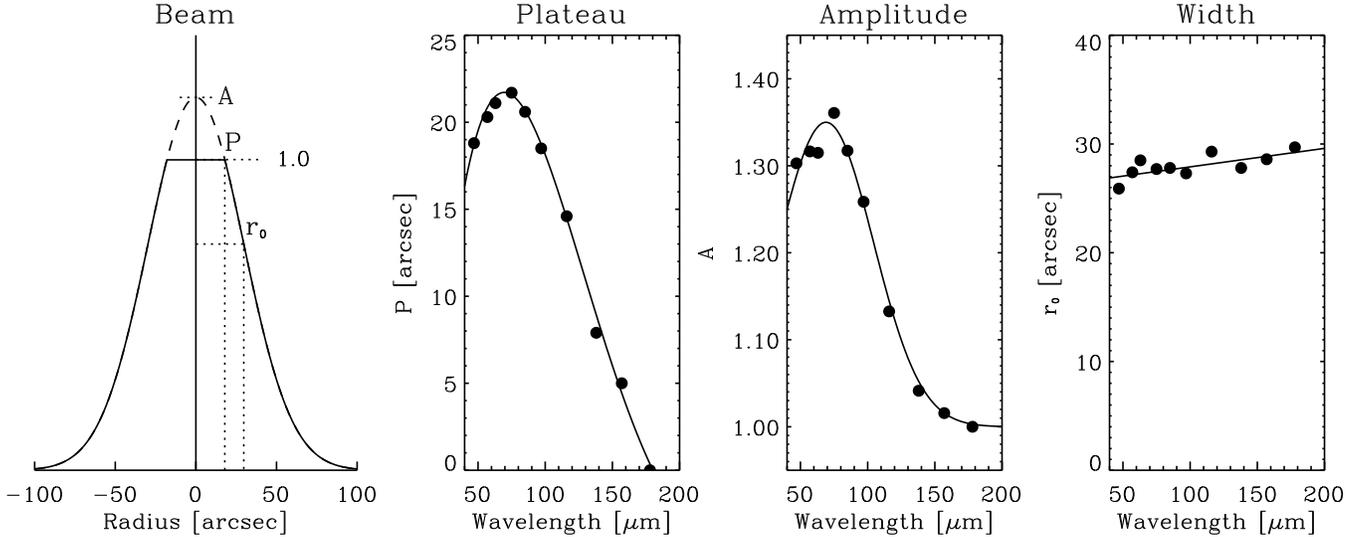}}
                        }
  \caption{The beam of the \lws\ at the centre wavelengths of its ten detectors (SW\,1 to SW\,5
  and LW\,1 to LW\,5, see Clegg et al. 1996). Azimuthal averages are represented by the filled dots and the full lines
  refer to the continuous functions fitted to these data. For the distance $r$ from the beam centre
  (optical axis) in arcsec we have $\Omega_{\lambda}(r) = 1$ if $r < P$ and 
  $\Omega_{\lambda}(r) = A\,\exp [- \frac{1}{2}\,(\frac{r}{r_0})^2 ]$ if $r > P$, where
  the {\it plateau} $P$ is shown in the second panel from the left, the {\it amplitude} $A$ in the
  following panel and the {\it width} $r_0$ in the right hand panel. An explanatory sketch of this 
  parameterisation is provided in the farmost left hand panel}
  \label{lws_beam_size_fit}
\end{figure*}

A first order correction for the finite beam sizes to the flux density
(shown as small open symbols in Fig.\,\ref{smm1modbbfit}) was included by 

\begin{equation}
F_{\lambda} = \frac{\Omega_{\lambda,\,\rm LWS}}{\Omega_{\lambda,\,B}}\left(\frac{\Omega_{\lambda,\,S} +
\Omega_{\lambda,\,B}}{\Omega_{\lambda,\,S} + \Omega_{\lambda,\,\rm LWS}}\right)\, F_{\lambda,\,B}
\end{equation}

where $\Omega_{\lambda,\,\rm LWS}$ is the beam size of the \lws\ at different wavelengths, $\lambda$ 
(see Section \ref{deconv} and Fig.\,\ref{lws_beam_size_fit}), $\Omega_{\lambda,\,B}$ is the beam size 
of the telescope in question, $\Omega_{\lambda,\,S}$ is the source size, and $F_{\lambda,\,B}$ is the 
source flux reported in the literature. Any dependence of the source size on the wavelength is
naturally accounted for by the radiative transfer modelling of Sect.\,\ref{rad_trans}. Here, it 
is assumed that the source size is the same as that of the blackbody fit and constant with wavelength.

This modified blackbody remains optically thick 
far beyond 100\,\um, up to $\lambda (\tau = 1) = 250$\,\um, the wavelength of 
unit optical depth. Longward of this wavelength the emission is optically thin 
and the total column density, $N$(\molh), can be estimated from

\begin{equation}
\tau_{\lambda} = \int_{}^{} \kappa_{\lambda}\, \rho(\ell)\,d\ell =
\kappa_{\lambda}\,\mu\,m_{\rm H}\,N({\rm H}_2)\,\left 
(\frac{M_{\rm gas}}{M_{\rm dust}} \right )^{-1}
\end{equation}

so that 

\begin{equation}
N({\rm H}_2) = \int_{}^{} n_{\rm H_2}(\ell)\,d\ell =
\frac{ \tau_{ \lambda_{0} } }{ \kappa_{ \lambda_{0} }\,\mu\,
m_{\rm H} }\, \frac{M_{\rm gas}}{M_{\rm dust}}
\end{equation}

\begin{figure}
  \resizebox{\hsize}{!}{
  \rotatebox{90}{\includegraphics{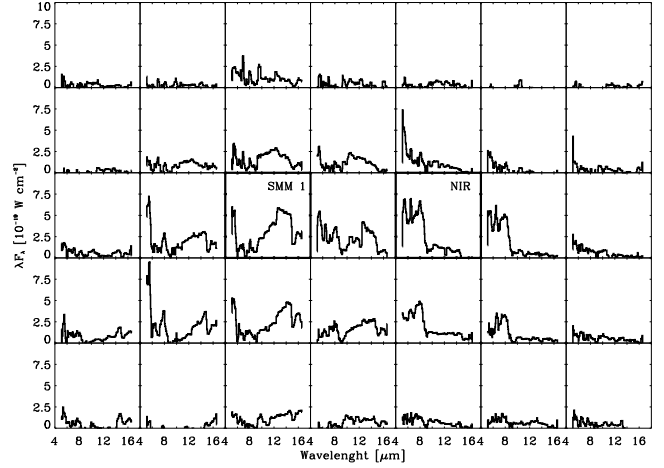}}
                        }
  \caption{\cvf\ view of the 5 to 15\,\um\ SED towards SMM\,1 inside the \lws\ beam. The pixel size is
  6\asec, hence the shown map is 42\asec\,$\times$\,30\asec. The map orientation is offset from
  the axes of the equatorial sysrem, rotated at a negative position angle of a few degrees. 
  Note the clear difference 
  between the SED of the \nir\ object (distinctly `blue') and that of SMM\,1 (distinctly `red') close to the 
  map centre and separated by less than 20\asec}
  \label{smm1_cvf}
\end{figure}

Taking $\kappa_{\lambda_0} = 1\,{\rm cm^{2}\,g^{-1}}$ at $\lambda_0 = 1.3$\,mm 
(Ossenkopf \& Henning 1994) and $\tau_{\lambda_{0}} = 0.05$, $\mu = 2.4$ for molecular gas 
and the gas-to-dust mass ratio $M_{\rm gas}/M_{\rm dust}$\,=\,\powten{2}, the average column 
density is $N$(\molh)\,=\,1.3\,$10^{24}$\,\cmtwo. For the assumed source radius of 1100\,AU, 
the mass of the optically thin envelope of SMM\,1 is 2.2\,$D_{310}^2$\,\msun. 
The average volume density $n$(\molh) corresponds then to several times \powten{7}\,\cmthree,
which is consistent with our choice of the theoretical opacities for grains with thin ice mantles for 
dense protostellar regions. However, the observationally determined value of $\beta$
seems only marginally compatible with the opacity law, for which $\beta = 1.8$. If one
accepts that the opacities are generally correct per se, the observed low $\beta$ values 
(see also Table\,\ref{tab_smm_bb_par}) could indicate that the optical depth through the
dust has been under-estimated even at long wavelengths.

The assumption of an isothermal, homogeneous sphere is of course very crude and physically
not at all compelling. However, the \lws\ data do not contain any information which can 
constrain the structure of the source. Hogerheijde et al. (1999), on the other hand, 
found evidence for structure in the visibility curves of their mm-wave data and adopted 
a radial power law distribution for the density, $n(r) \propto r^p$ with $p \sim -2$, 
so that their model of SMM\,1 is more centrally condensed. 
For a spherical, centrally heated optically thin dust envelope, a power law is also 
expected for the temperature profile, viz. $T(r) \propto r^q$. The exponent is 
in this case given by $q = - 2/(4 + \beta)$, where $\beta$ refers as before to the 
wavelength dependence of the absorption cross section of the dust grains. 
Since $\beta_{\rm obs} \sim 1$, the temperature profile could be expected to
exhibit $q \sim -0.4$, a value which was indeed also used by Hogerheijde et al. (1999). 

It has been suggested that SMM\,1 is a so called Class\,0 source (Andr\'e et al. 1993)
and the correct assessment of its envelope mass (and central mass, of course) is of 
particular importance for such type of object. In the following section, we shall attempt to 
improve on the reliability of the parameters for SMM\,1 by treating the radiative transfer 
through the dust envelope in a self-consistent manner.

Before doing that, we need to address the possible near-infrared excess over
a Class\,0 SED, however (see Fig.\,\ref{smm1modbbfit}). It is clear that
over the 7\asec\ source size, the deduced column density, 
$N$(\molh)\,=\,\powten{24}\,\cmtwo, would imply an 
accompanying visual extinction of \av\,\about\,\powten{3}\,mag
(Bohlin et al. 1978, Savage \& Mathis 1979), or still nearly $A_{\rm K} \sim 100$\,mag at 2\,\um\ 
(Rieke \& Lebofsky 1985). Therefore, on arcsecond scales around SMM\,1 one would hardly 
expect any detectable direct flux in the near-infrared. 
This assertion is confirmed by \isocam-\cvf\ observations (Table~\ref{obstab}) which show that 
the \nir\ flux originates from a source different from SMM\,1. In Fig.\,\ref{smm1_cvf}, a map of
the spectra on the 6\asec\ pixel scale are shown. Obviously, SMM\,1 and the source west
of it do not have the same SEDs in the 3 to 15\,\um\ region. Hence, we conclude that the SED of SMM\,1 
contains less \nir\ flux than indicated in Fig.\,\ref{smm1modbbfit} and, consequently,
these observations are supporting the view that SMM\,1 is indeed a Class\,0 object.

\begin{figure}
  \resizebox{\hsize}{!}{
  \rotatebox{90}{\includegraphics{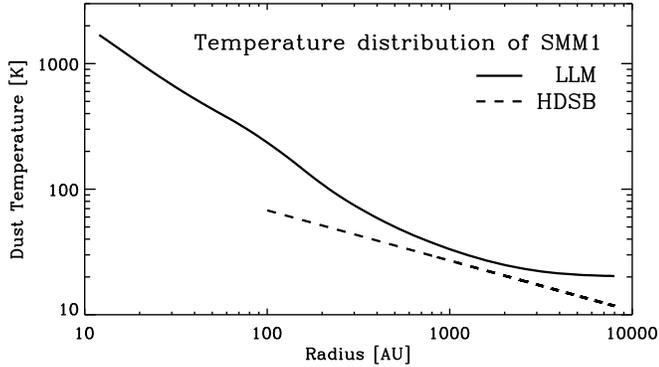}}
                        }
  \caption{The radial run of the temperature in the (assumed) spherical envelope of SMM\,1. The solid
  line shows the optically thick dust envelope, obtained from the self-consistent radiative transfer model 
  discussed in the text (LLM = this paper). For comparison, the optically thin model with $T \propto r^{-0.4}$
  by Hogerheijde et al. (1999), and identified as HDSB, is shown by the dashed line. Both models assume an 
  $r^{-2}$ density law and the same grain opacities. When compared to the optically thin 
  assumption, the run of the temperature is markedly different: a considerably steeper gradient
  is needed in most regions of the envelope to drive out the centrally generated flux (\about\,70 \lsun)}
  \label{smm1_temp}
\end{figure}

\begin{figure}
  \resizebox{\hsize}{!}{
  \rotatebox{90}{\includegraphics{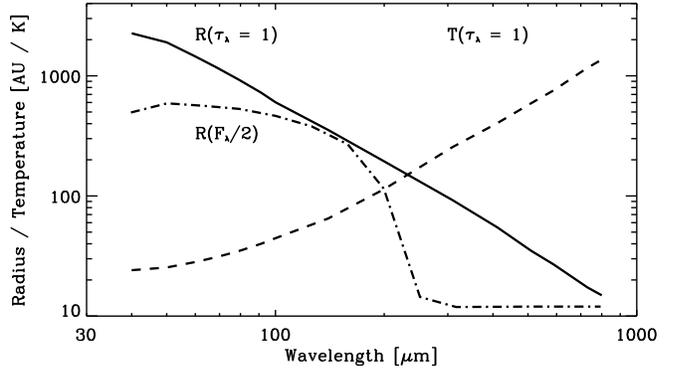}}
                        }
  \caption{Surfaces of unit optical depth of the model of SMM\,1 in the far infrared and sub-millimetre (40\,\um\ to
  about 1\,mm). Shown is the variation of the source size with wavelength, i.e. the radius, $R(\tau_{\lambda}=1)$ in AU, 
  and the corresponding dust temperature at this radial distance from the central heating source, $T(\tau_{\lambda}=1)$ in K.
  The dash-dotted line, labelled $R(F_{\lambda}/2)$, designates the locus of the radius at which half of the flux at a given
  wavelength is emitted}
  \label{smm1_temp_rad}
\end{figure}

\subsubsection{Self-consistent radiative transfer models of SMM\,1 \label{rad_trans}}

The models by Hogerheijde et al. (1999) are based on the assumption of optically thin emission. 
However, it was realised in the previous section that optical depth effects might affect the SED
of SMM\,1 even at long wavelengths. In order to test this hypothesis we have run models of the
transfer of radiation through the dust envelope of the source. A fuller description of these 
computations will be presented in a forthcoming paper (Larsson et al. in prep.). From the comparison 
with the observations, these self-consistent models provide the temperature distribution in the 
envelope. For instance, Fig.\,\ref{smm1_temp} depicts the true temperature profile of the 
Hogerheijde et al. model of SMM\,1 (see below). For the general cloud background we estimate
a temperature of 20\,K, and that flux has been subtracted from the model of SMM\,1.

Complying with Hogerheijde et al. (1999), we adopt a spherically symmetric geometry of the 
dust envelope about a central heating source and a radial power law distribution for the density, 
with $p = -2$ (in order to reproduce the visibilites obtained by these authors), i.e. 
$n(r)=1.7\times 10^{11}\,(r/12\,{\rm AU})^{-2}$\,\cmthree, which reproduces the density at 100\,AU 
we inferred from that paper. As before,the opacities were taken from Ossenkopf \& Henning (1994), 
specifically the model identified as
{\it MRN with thin ice mantles, density} \powten{7}\,\cmthree\ {\it and age} \powten{5}\,yr  
(although one would not expect these thin ice mantles to survive in the central regions of 
the envelope, where temperatures are greately in excess of 100\,K, we kept this particular choice 
of opacities for comparison with the other models; using other grain compositions did not 
change the {\it general} results). 

\begin{table*}
\caption{\label{tab_smm_bb_par} Dust temperatures, sizes and opacity indices of IR/SMM envelopes in the \scc}
 \resizebox{\hsize}{!}{
  \begin{tabular}{l cccr cccr cccr l} 
  \hline
Source Name  	 & \multicolumn{3}{c}{$T_{\rm dust}$ (K)} & 
                 & \multicolumn{3}{c}{$\oslash$  (arcsec)} &
                 & \multicolumn{3}{c}{$\beta_{\rm obs}$} &
                 & Note$^{\P}$ \\
\cline{2-4} \cline{6-8} \cline{10-12}
                 & bb$^{\dagger}$   & $\tau_{100\mu\rm m}$=1$^{\dagger}$   & {\sc hb}  & 
                 & bb$^{\dagger}$   & $\tau_{100\mu\rm m}$=1$^{\dagger}$   & {\sc hb}  &
                 & {\sc llm}$^{\S}$ & {\sc dmrdr}                          & {\sc ced} &
                 &        \\
  \noalign{\smallskip}
  \hline
  \noalign{\smallskip}
SMM\,1/FIRS\,1         & 36\, & 43 & 27 &
                       & $7 \pm 1$\, & 4 & 12 & 
                       & $1.2 \pm 0.2$ & 0.9 & $0.5 \pm 0.2$  & 
                       &  \\
SMM\,2                 & 33:  & 36 & 24 &
                       & $3 \pm 2$: & 1.5 & 20 & 
                       & $0.2 \pm 1.9$ & 1.1 & $0.8 \pm 0.8$ & 
                       & 1 \\
SMM\,3                 & 31\, & 32 & 24 &
                       & $5 \pm 2$\, & 3 & 20 & 
                       & $0.6 \pm 0.2$ & 0.8 & $-0.1 \pm 0.4$\,\,\,\,   & 
                       &  \\
SMM\,4                 & 30\, & 27 & 20 &
                       & $5 \pm 2$\, & 4 & 20  & 
                       & $0.8 \pm 0.2$ & 0.8 & $0.5 \pm 0.3$ &  
                       &  \\
SMM\,5/EC\,53          & & & &
                       &     &    & 20  &
                       & $0.9 \pm 0.9$ & 0.9 & $1.6 \pm 0.5$ & 
                       &  1,\,2,\,3 \\   
SMM\,6/SVS\,20         & & & &   
                       &     &    & 20  & 
                       & $0.5 \pm 2.4$ & 0.8 & $1.3 \pm 1.0$ & 
                       &  1,\,2,\,3  \\   
SMM\,8                 & 21: & &  &   
                       & $6 \pm 3$:    &     &  & 
                       & & & & 
                       &  \\   
SMM\,9/S\,68N          & 24\, & 27 & 23 &
                       & $9 \pm 4$\, & 3.5 & 20  & 
                       &  & 1.2 &  & 
                       &         \\      
SMM\,11/HB\,1          & 31: & & &
                       & $4 \pm 2$: &  &  & 
                       & & & & 
                       &  1,\,3    \\   
HB\,2                  & 24: & & &
                       & $5 \pm 3$: & &  & 
                       & & & & 
                       &   \\   
SVS\,2                 & 25: & & &
                       & $9 \pm 5$: &  &  & 
                       & & & & 
                       & 1,\,3 \\                      
  \noalign{\smallskip}
  \hline
  \noalign{\smallskip}  
  \end{tabular}
}   
References: {\sc hb} = Hurt \& Barsony (1996), {\sc llm} = this paper, {\sc dmrdr} = Davis et al. (1999), 
{\sc ced} = Casali et al. (1993). \\
Notes to the table: \\
$^{\P}$ 1 = extended emission, 2 = excluded from calculation, 3 = near infrared source (see the text). \\
$^{\dagger}$ Temperature and diameter values are for bb = blackbody fit
and $\tau_{100\mu\rm m}=1$ refers to the radiative transfer model. \\
$^{\S}$ Our $\beta_{\rm obs}$ has been evaluated between 0.8\,mm and 3.4\,mm.  \\
\end{table*}

The model reproduces indeed the visibilites of SMM\,1. Further, the observed SED is reasonably well fit
at long wavelengths (Fig.\,\ref{smm1modbbfit}), i.e. the observed $\beta$-value is obviously reproduced.
This is mainly because the envelope is optically thick out to nearly 
1\,mm ($\lambda_{\tau=1} = 900$\,\um, see the Tables \ref{tab_smm_bb_par} and
\ref{tab_smm_phys_param}, where also other model results are provided). As expected, the 
``blackbody parameters'' of the model are not vastly different from the single temperature fit: 
e.g., the diameter and temperature of the spherical source are 4\asec\ and 43\,K, respectively,
for unit optical depth at 100\,\um, viz. for $\tau_{100\,\mu{\rm m}} = 1$.  
Values at other 
wavelengths are presented graphically in Fig.\,\ref{smm1_temp_rad}. Evidently, at the longer
wavelengths, one `looks' progressively deeper into the envelope, towards hotter and denser regions.
Further, at the longer wavelenghts more than half of the flux originates from the inner regions, 
where the dust destruction front is located. This is shown in the figure as $R(F_{\lambda}/2)$, i.e.
the radius at which half of the flux at a given wavelength is emitted.

\begin{table*}
\caption{\label{tab_smm_phys_param} Luminosities, opacities and masses of SMM envelopes in the \scc}
\resizebox{\hsize}{!}{
  \begin{tabular}{l cccr cc  c cccc l} 
  \hline
Source Name 	 & \multicolumn{3}{c}{$L^{\dagger}$ (\lsun)} &
                 & \multicolumn{2}{c}{$\lambda_{\tau=1}^{\pounds}$ (\um)}
                 & $N$(\molh)$^{\ddagger}$
                 & \multicolumn{4}{c}{$M_{\rm envelope}^{\dagger}$ (\msun)}
                 & Note \\
     \cline{2-4} \cline{6-7} \cline{9-12}
                 & {\sc llm} & {\sc hb} & {\sc ced}  & 
                 & {\sc i}   & {\sc ii} 
                 & (\powten{24}\,\cmtwo)   
                 & {\sc i}   & {\sc ii} & {\sc hb} & {\sc hdsb}  
                 &        \\
  \noalign{\smallskip}
  \hline
  \noalign{\smallskip}
SMM\,1/FIRS\,1  & 71 & 46 &     84      & & 250 &  900 &  1.3  & 2.2 & 6   & 3   & 5.2 &     \\
SMM\,2          & 10 &  6 & $0.8 - 32$  & & 170 &  970 &  0.7  & 0.3 & 2.3 & 0.6 &     & Extended \\
SMM\,3          & 18 &  8 &   $2 - 32$  & & 170 & 1000 &  0.7  & 0.6 & 4   & 0.9 & 1.8 &     \\
SMM\,4          & 16 &  9 &   $3 - 32$  & & 270 & 1400 &  1.5  & 1.2 & 6   & 3   & 3.2 &     \\
SMM\,9/S\,68N   & 16 &  6 &             & & 160 & 1200 &  0.6  & 1.7 & 5   & 1.0 &     &        \\
  \noalign{\smallskip}
  \hline
  \noalign{\smallskip}  
  \end{tabular}
}
References: {\sc llm} = This paper, {\sc hb} = Hurt \& Barsony (1996), {\sc ced} = Casali et al. (1993), 
{\sc i} = Single temperature/optically thin approximation, {\sc ii} = Radiative transfer model, 
{\sc hdsb} = Hogerheijde et al. (1999). \\
Notes to the table: \\
$^{\dagger}$ All reduced to the same distance, viz. $d = 310$\,pc. Our luminosities refer to 40 -- 3\,000\,\um. \\
$^{\pounds}$ Opacities from Ossenkopf \& Henning (1994: MRN with thin ice mantles, \powten{7}\,\cmthree\ and
\powten{5}\,yr, for which $\beta = 1.8$). \\
$^{\ddagger}$ Refers to {\sc i} (single temperature, optically thin approximation).
\end{table*}

We shall not dwell further on the details of these specific radiative transfer calculations,  
for the following reason: although the model fit is quite acceptable in the FIR and submm/mm spectral 
region, it clearly underproduces the flux at wavelengths shortward of the peak of the SED 
(Fig.\,\ref{smm1modbbfit}). The remedy of this shortcoming requires the relaxation of the 
assumption regarding the source geometry. Because of disk formation, spherical geometry is generally 
not considered a good approximation for YSOs, including protostellar objects. It is possible, in fact, 
to satisfactorily fit the entire SED using a radiative transfer model in a 2d axially symmetric geometry 
(disk plus envelope). The results of such computations will be presented in a follow-up paper 
(Larsson et al. in prep.).

\begin{figure}
  \resizebox{\hsize}{!}{
  \rotatebox{90}{\includegraphics{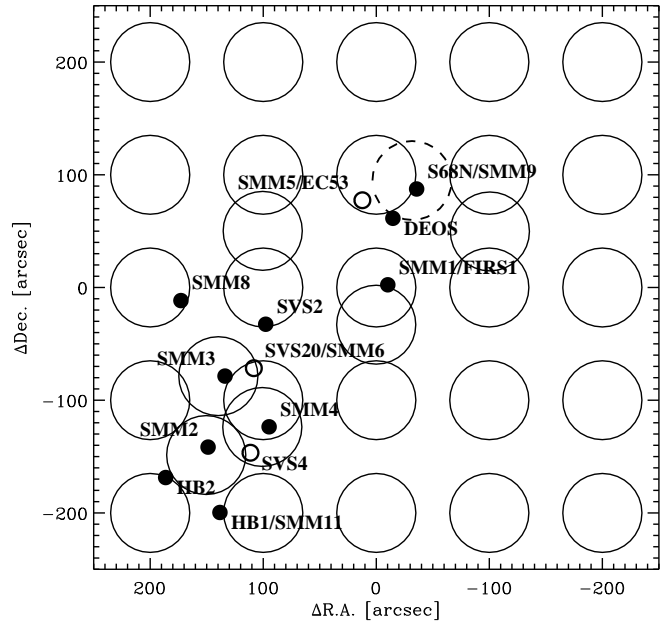}}
                        }
  \caption{The positions observed with the \lws\ and those of the sources included in the 
  {\it Maximum Likelihood Method} deconvolution. Solid circles correspond to the \lws\ 
  beams and the dashed circle especially marks an \lws\ observation, which is taken at a different
  time than the grid points in the map (see Sect\,\ref{nwsource}). The filled dots refer 
  to the \fir/sub-mm sources for which the deconvolution computations were performed - 
  in the wavelength range 50 to 200\,\um\ -  whereas the small open rings identify sources 
  which were omitted in the final \fir\ point source modelling (see the text)}
  \label{lucy_sources}
\end{figure}

\subsection{The other SMMs of the Serpens cloud core}

In Fig.\,\ref{lwscontmap}, a secondary emission peak  
lies on the extension towards the south-east.
This morphology is very similar to that previously seen at \fir\ wavelengths
(e.g., Nordh et al. 1982, Harvey et al. 1984, Hurt \& Barsony 1996)  
and also at longer wavelengths (Casali et al. 1993, Testi \& Sargent 1998, 
Davis et al. 1999). In particular at higher spatial resolution, this emission region
breakes up into numerous point-like sources. Because of this source crowding, 
severe source confusion within the \lws-beam can be expected at \fir\ wavelengths. 
In the next section, we shall attempt to disentangle the various contributions 
from the different sources. This is possible, because spatially filling-in and 
overlapping \lws\ data are in existance (see Table~\ref{obstab} and 
Fig.\,\ref{lucy_sources}). 

\subsubsection{SEDs of confused sources: Spectro-spatial deconvolution of the \lws\ data \label{deconv}}

\begin{figure*}
  \resizebox{\hsize}{!}{
  \rotatebox{90}{\includegraphics{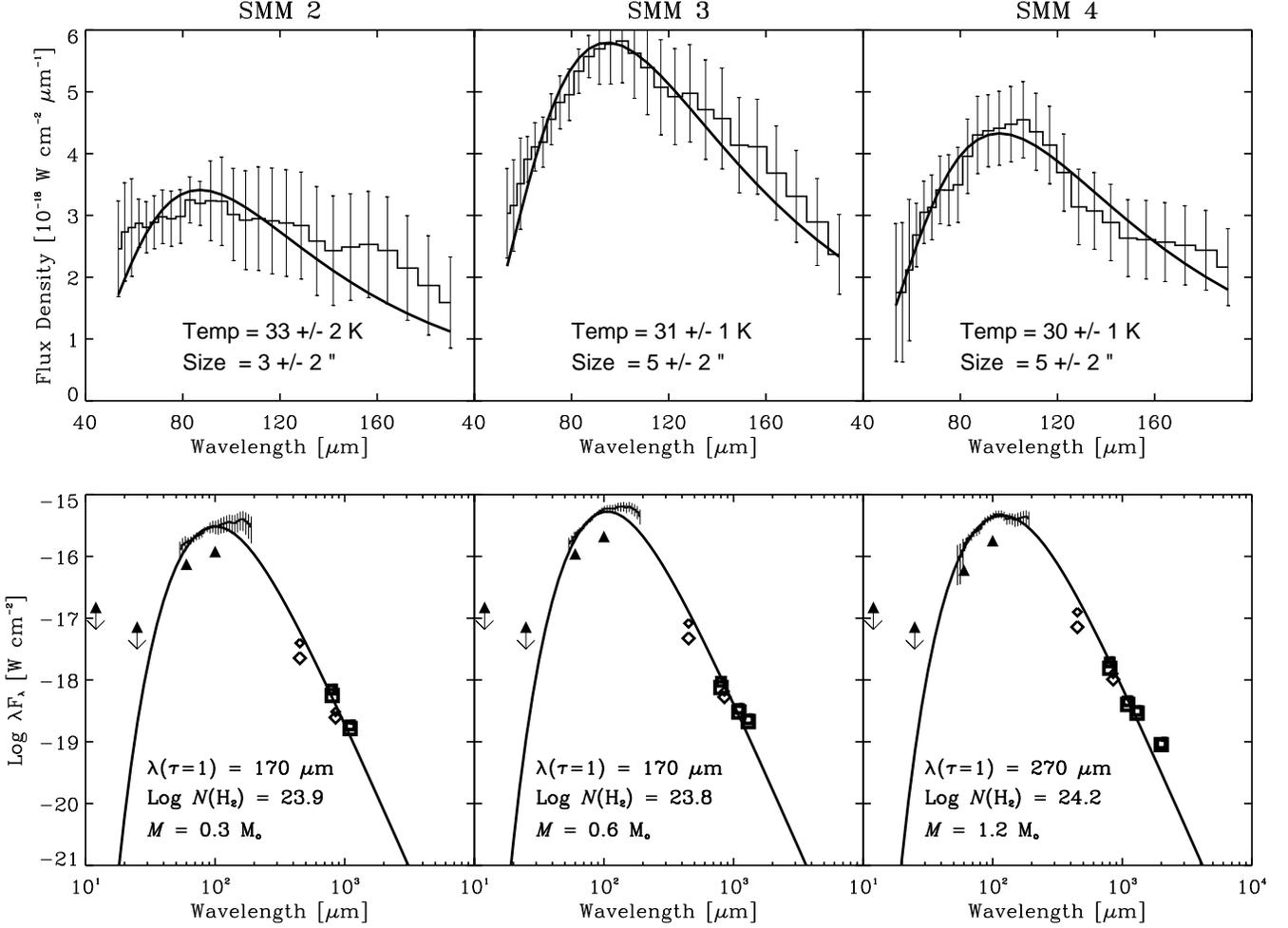}}
                        }
  \caption{ {\bf Upper panels:} The \lws\ spectra of SMM\,2, SMM\,3 and SMM\,4, which 
  are the results of the spectro-spatial deconvolution technique described in the text. 
  The solid lines illustrate the blackbody fits to the data which are shown as histograms. 
  The error-bars are given by (cf. Appendix A):
  $\Delta F_S = \sum \Omega\,\Delta F_M/ \sum \Omega$, where
  $\Delta F_M = \Delta F_O + \mid\! F_O - \sum \Omega\,F_S\!\mid$ and where
  $\Delta F_O$ are the errors displayed in Fig.\,\ref{rawdata}.
  {\bf Lower panels:} Modified blackbody fits over the full SED range (10 -- \powten{3}\,\um). 
  Literature and/or \iso\ data are coded as in Fig.\,\ref{smm1modbbfit} }
  \label{mlm_2_3_4}
\end{figure*}

To search for the \fir\ sources and to deduce their associated SEDs 
we developed a program for a {\it Maximum Likelihood Method} (e.g., Lucy 1974)
and applied it to the  entire set of \lws-map data (Table~\ref{obstab}). 

The first of two basic assumptions is that all sources are point-like, 
which is probably reasonable considering the large beam size of the \lws\ 
(\hpbw\,\about\,70\asec). In support of this, SMM\,1 was found, in the previous 
sections, to be a point source. Secondly, the exact positions of all sources
are assumed known and kept fixed. These were taken from \isocam\ observations 
in the mid-infrared (Kaas 1999) and/or from observations in the sub-millimetre 
by Davis et al. (1999). We then pick an arbitary grid point within the field of the 
\lws-map and try to estimate, at that location and for a fixed frequency, 
the radiative flux which originated from all sources in the field. 
This is done at all locations and for all \lws\ wavelengths in an iterative scheme as
described in more detail in Appendix\,A.

For the restauration of the source spectra, the band-width corresponded always 
to the spectral resolution $R_{\lambda}=20$. For the wavelength dependent beam pattern 
of the \lws, $\Omega_{m,s}$, the azimuthal averages of the presently best known 
values were used (see Fig.\,\ref{lws_beam_size_fit}). This was done for all ten detectors of the \lws\ 
and out to \about\,150\asec\ from the beam centre, which would correspond to about the 5\% level 
of a perfectly Gaussian beam. The observational basis for this was raster mapping 
of the point-like planet Mars (B.M.\,Swinyard and C.\,Lloyd 1999, private communication). 

\begin{figure*}
  \resizebox{\hsize}{!}{
  \rotatebox{90}{\includegraphics{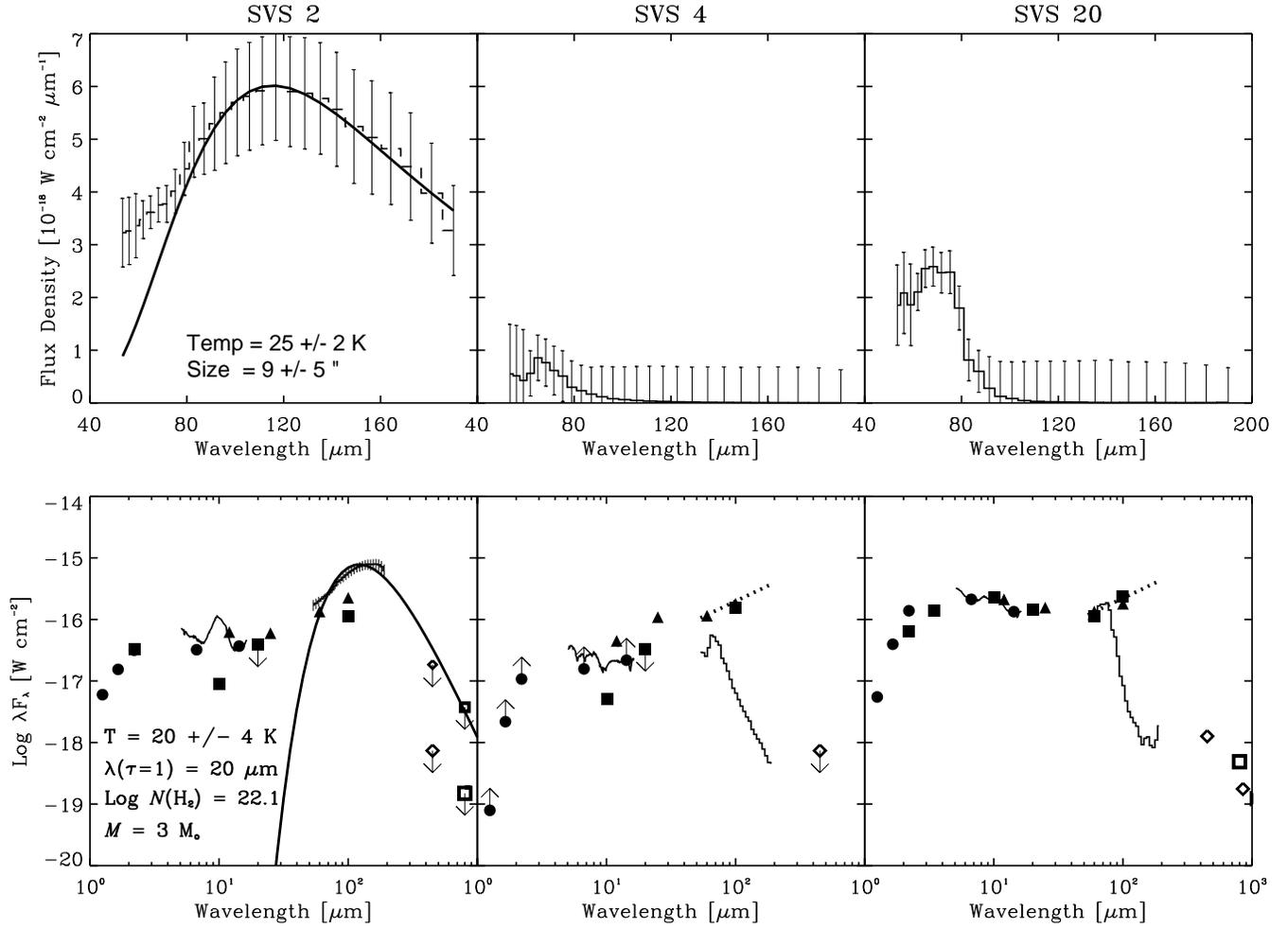}}
                        }
  \caption{ Same as Fig.\,\ref{mlm_2_3_4} but for SVS\,2, SVS\,4 and SMM\,6/SVS\,20. 
  SVS\,2 is actually better modelled as extended emission in the far-infrared. This is shown 
  in the panel below by the larger upper limits in the submillimetre, which refer to a source 
  filling the \lws\ beam. For SVS\,4 and SVS\,20, the dotted lines in the lower panels depict the 
  deduced $3\,\sigma$ upper limits on the flux $\lambda F_{\lambda}$ in the \lws\ regime }
  \label{mlm_svs2_4_20}
\end{figure*}

In Fig.\,\ref{lucy_sources}, the sources which were included in the modelling are identified
and shown together with the pointings of the \lws. These sources are 
recognized as the strongest ones at sub-millimetre (SMM\,$1 - 11$, except SMM\,7 and 10) and/or 
at mid-infrared wavelengths (SVS\,2, SVS\,4 and SVS\,20). The spectra were corrected for
background emission, which was defined along the northern and western edges of the map (see
Figs.\,\ref{rawdata} and \ref{smm1bbfit}).
For these prominent sources, convergence was typically achieved after five iterations.
In general, the computations were halted after visual inspection of the results, but
tests were run up to 1000 iterations demonstrating that the solutions were stable. In 
Appendix\,B, a demonstration of the algorithm's reliability is provided using known source spectra. 
It is also gratifying to note that the procedure left, as could be expected, the spectrum 
of SMM\,1 essentially unaltered, i.e.
the {\it Maximum Likelihood Method} solution for this source is entirely within the quoted errors.
Yet another check was provided by an \lws\ Guaranteed Time observation half a beamwidth south
(labelled `Flow' in Table\,\ref{obstab}), which was correctly recovered by our method. Since
these observations were apart in time by 1\,year, this result also indicates that no
significant \fir-intensity variations have occurred on that timescale.

To properly appreciate the results to be presented below and summarised in
Tables \ref{tab_smm_bb_par} and \ref{tab_smm_phys_param}, it might be useful to 
remember that the method attempts
to retrieve information which is hidden inside the spatial resolution of the observations.
Therefore, these solutions represent, at best, very likely possibilities of what these
source spectra might look like. Note that the 30 spectral bins were treated as
independent observations, so that the fact that the resulting SEDs are continuous and have
a meaningful appearance, lends confidence in the method. There is no a priori reason that this should 
be the case (cf. Eqs.\,A1 to A9). However, confirmation must ultimately come from direct observation.
With this proviso in mind, the results of these deconvolution computations are examined
for individual sources in subsequent sections. 

\subsubsection{Spectro-spatial deconvolution results: the south-eastern sources}

\begin{figure*}
  \resizebox{\hsize}{!}{
  \rotatebox{90}{\includegraphics{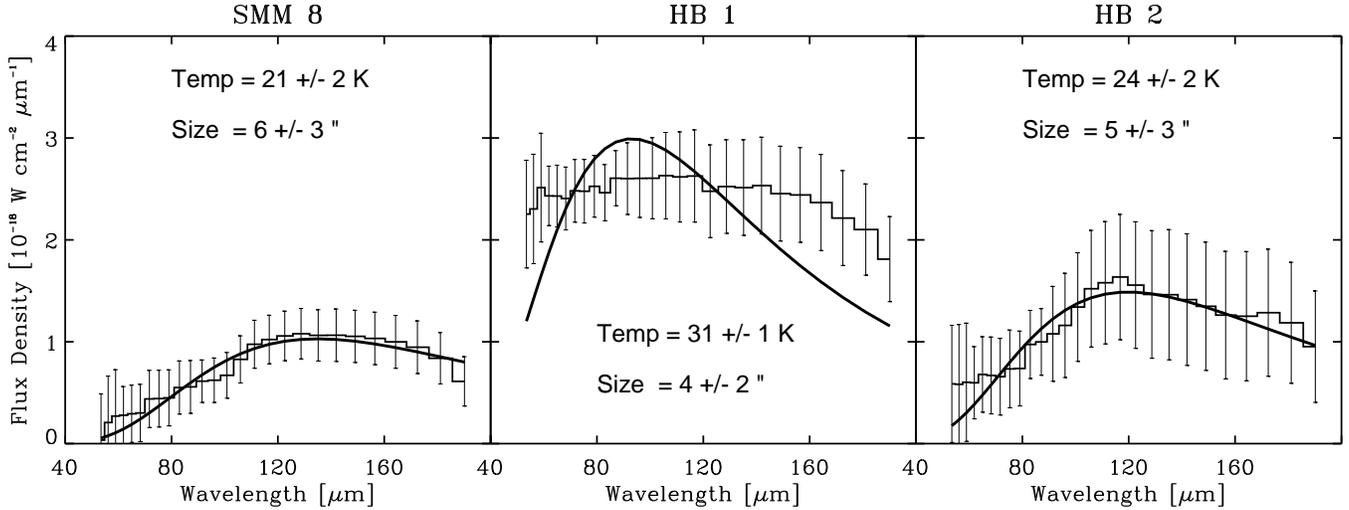}}
                        }
  \caption{Similar as Fig.\,\ref{mlm_2_3_4} but for SMM\,8, HB\,1 and HB\,2, except that only
  the upper panels have been included}
  \label{mlm_8_hb12}
\end{figure*}

In the south-east part of the map, the level of confusion is high,
with many sources contributing to the observed emission. A few sources are
however dominant, viz. in the near infrared SMM\,6/SVS\,20 and at  
submillimetre wavelengths SMM\,3 and SMM\,4, respectively.  

{\underline {SMM\,2, 3 and 4:}} Satisfactory solutions were obtained for the sources 
SMM\,3 and SMM\,4, whereas the broad spectrum of SMM\,2 cannot plausibly be represented 
by a modified single-temperature blackbody (see Fig.\,\ref{mlm_2_3_4}). This is consistent 
with the interferometric observations by Hogerheijde et al. (1999) and by Choi et al. (1999), 
which failed to reveal a central condensation for SMM\,2. Kaas (1999) found a $K$\,-\,band
nebulosity at this position. We interpret its \lws\ spectrum therefore to be that of an extended structure. 

Dust temperatures of SMM\,2 -- 4 are of the order of 30\,K (Table\,\ref{tab_smm_bb_par}),
which is in general agreement with earlier work (Harvey et al. 1984), but which again contrasts
to the lower values obtained by Hurt \& Barsony (1996). Also the extent of the \fir\ photospheres
is considerably smaller, on the order of a few arcsec rather than ten arcsec. For these optically
thick envelopes (Tab.\,\ref{tab_smm_bb_par}) the masses, which are in the range 2 to 6\,\msun, would 
be severly underestimated by the optically thin approximations (Tab.\,\ref{tab_smm_phys_param}).

{\underline {SVS\,2, SVS\,4 and SMM\,6/SVS\,20:}} The point source assumption leads immediately 
to inconsistencies for SVS\,2, which would become unreasonably bright in the far infrared,
which is highly unlikely given the non-detection of the source at longer wavelengths 
(Fig.\,\ref{mlm_svs2_4_20}).
An overall better solution is obtained for an extended source, filling the \lws\ beam. 
The modelling by Hurt \& Barsony (1996) gave a very different picture, which yielded three
nearly equally strong \fir\ point sources, with SVS\,2 being only slightly stronger at 100\,\um. 

\begin{figure*}
  \resizebox{\hsize}{!}{
  \rotatebox{90}{\includegraphics{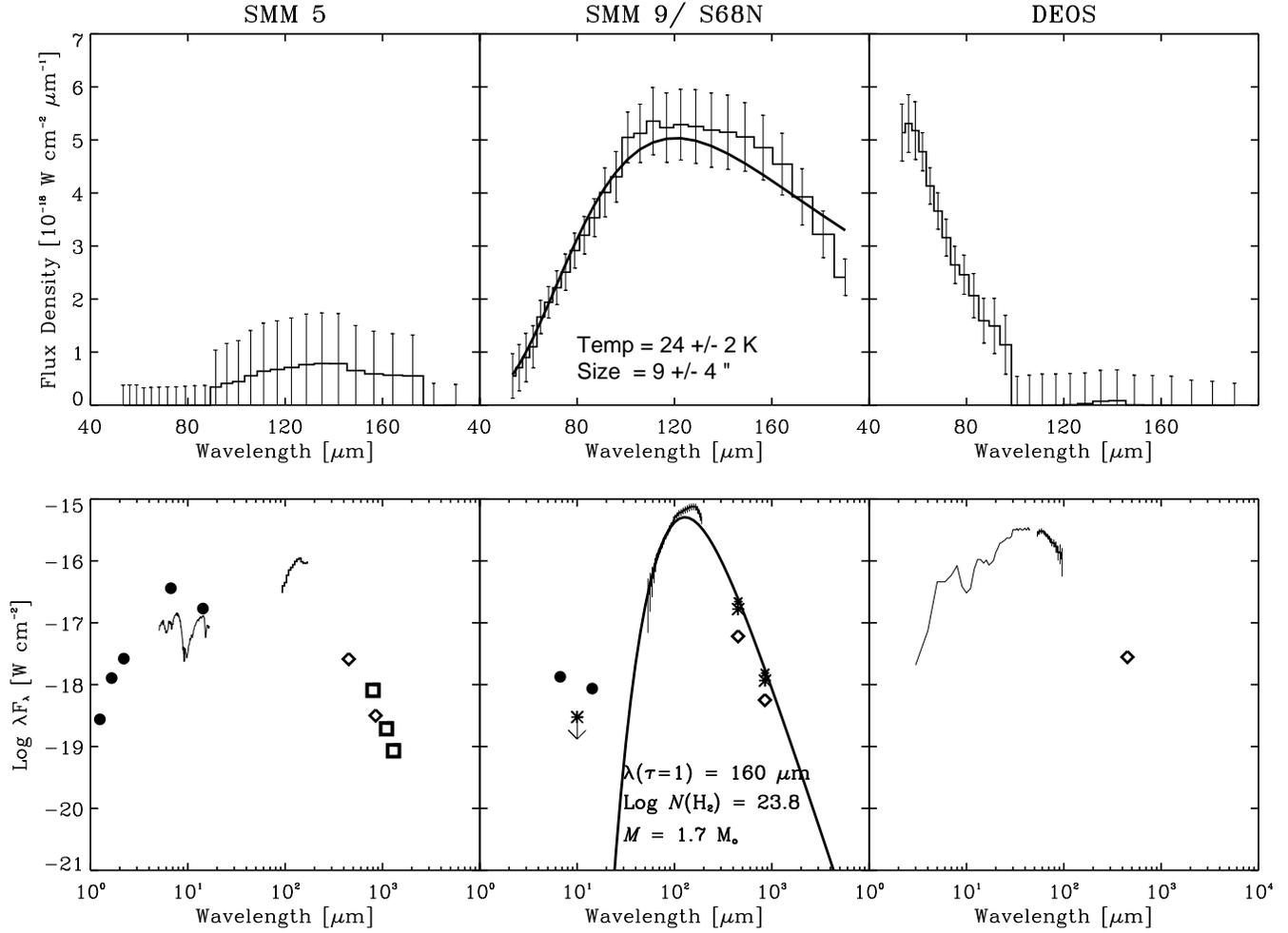}}
                        }
  \caption{Same as Fig.\,\ref{mlm_2_3_4} but for SMM\,5, SMM\,9/S\,68N and DEOS. The data shown 
  by the asterisks in the SED of SMM\,9/S\,68N  are from Wolf-Chase et al. (1998)}
  \label{mlm_5_9_deos}
\end{figure*}

The modelling results in largely undetectable \lws\ fluxes for SVS\,4 and SVS\,20. 
In particular for SVS\,4, these low flux levels are consistent with the lack of 
detection at 3\,mm, whereas for SVS\,20 some emission is possibly present
at the shortest \lws\ wavelengths. The objects of this group are clearly visible at near infrared
wavelengths, as stellar clusters and nebulosity and in Table\,\ref{tab_smm_bb_par}, 
these objects are therefore identified as \nir\ sources. Further, our solutions for the \lws\ data 
provide reasonable complements to the \cvf\ data (see Fig.\,\ref{mlm_svs2_4_20}), 
which have been obtained at the considerably higher spatial resolution of 6\asec\ and, hence, 
are as such not dependent on any models.  

{\underline {SMM\,8, SMM\,11/HB\,1 and HB\,2:}} These sources represent the fainter members of the 
south-eastern group in the \scc. The objects designated PS (Point Source) by Hurt \& Barsony (1996)
are labelled HB\,1 and HB\,2 in Figs.\,\ref{lucy_sources} and \,\ref{mlm_8_hb12}. 
Whereas the algorithm finds reasonable solutions
for SMM\,8 and HB\,2, identifying them as point sources of roughly 5\asec\ size at $T$\,\gapprox\,20\,K,
it fails for SMM\,11/HB\,1. At that location, Kaas (1999) recently found a source at 2\,\um.

\subsubsection{Spectro-spatial deconvolution results: the north-western sources \label{nwsource}}

The members of this group, north of SMM\,1, include SMM\,9/S\,68N, SMM\,5 and 
DEOS, the `Deeply Embedded Outburst Star' of Hodapp et al. (1996). As previously
pointed out, the strong \fir\ source SMM\,1 is comfortably displaced 
from the other three sources, the \lws\ data of which are quite severely confused
(see Fig.\,\ref{lucy_sources}). These confusion problems have also been encountered 
by Hurt \& Barsony (1996), who divided the total flux into three equal parts among 
these objects. 

{\underline {SMM\,5 and SMM\,9/S\,68N:}} At the position of SMM\,5, no 
recognisable point source was found at \lws\ wavelengths (Fig.\,\ref{mlm_5_9_deos}). The object
coincides with a bow-shock shaped nebula and it could be that 
no well defined pre-stellar condensation exists. On the other hand, it is quite likely that
SMM\,5 simply is relatively too weak to be detectable. In order not to unnecessarily increase
the noise, the source was excluded from further modelling. 

Towards the S\,68N complex, two independent \lws\ observations exist: one
is a pointed observation, the other a grid point in our map. Spatially, these
two observations are separated by about half a beamwidth (33\arcsec), and 
temporally, by about half a year (Table~\ref{obstab}). 
The difference in flux amounts to about a factor of two, as would be expected, 
if S\,68N solely would dominate the \fir\ emission in this region. 
The observations by Davis et al. (1999) revealed SMM\,9/S\,68N to be
the dominant submm source, by factors of about 2 -- 3 in flux. We expect, therefore, this
source to be the brightest also in the far infrared, i.e. at the {\it longest} \lws\ wavelengths.
This seems indeed to be the case, as shown in Fig.\,\ref{mlm_5_9_deos}, where the 
modelling reveals a relatively cool source ($T$\,\gapprox\,20\,K). 

{\underline {DEOS:}} A further underlying assumption of our
deconvolution method is that the emission does not vary with time. This was previously 
taken tacitly for granted, since this condition is normally fulfilled in the far infrared.
For the outburst source DEOS, there is a potential risk, however, that this may not apply.
Between August 1994 and July 1995, this object had increased its brightness by 
almost 5\,mag in the $K$-band (Hodapp et al. 1996). 

\begin{figure}
  \resizebox{\hsize}{!}{\rotatebox{90}{\includegraphics{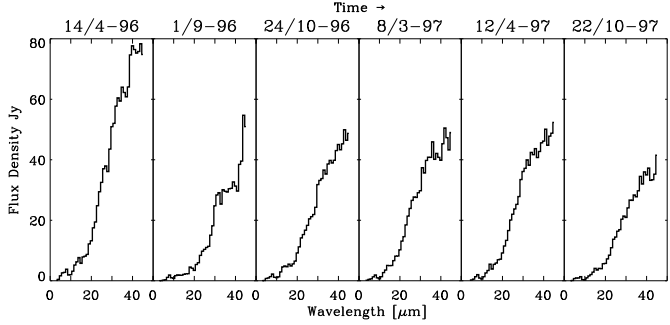}}}
  \caption{Time evolution of the shortwave SED of DEOS, obtained with the \sws\ in the
  wavelength interval 2 -- 45\,\um. After April 14, 1996, the spectral flux density had 
  decreased by a factor of about two and then stayed constant until the last measurement,
  i.e. October 22, 1997}
  \label{swsfuor}
\end{figure}

At near to midinfrared wavelengths, DEOS was observed, during about one year and a half, 
by T.\,Prusti with the \sws\ on six different dates
(see Table\,\ref{obstab}: April 14, 1996, to October 22, 1997). The time
evolution of the short-wave SED is displayed in Fig.\,\ref{swsfuor}, from which
it is evident that, between April and October 1996, the \sws\ flux had dropped by 
a factor of two. This is remarkably similar to what happened to the (quasi-)simultaneous
\lws\ spectra and led us to suspect that some or most of the \fir\ flux in fact
was due to DEOS rather than to SMM\,9, as might have been concluded on the basis
of the discussion in the previous paragraph.

\begin{table}
\begin{flushleft}
\caption{\label{tabcam} Aperture photometry in \isocam\ images in Jy/beam,
where the beam is that of the \sws}
  \begin{tabular}{llccc}
  \hline
  Object      & Filter   & 14 Apr 96        & 22 Sep 97        & Ratio \\
  \hline
  \noalign{\smallskip}
  
  DEOS        & LW\,2    & 2.8              & 2.2              & 1.27  \\
  DEOS        & LW\,3    & 6.8              & 5.4              & 1.26  \\
  \\
  EC\,53      & LW\,2    & 0.90             & 0.90             & 1.00  \\
  EC\,53      & LW\,3    & 1.12             & 1.15             & 0.97  \\
  \hline
  \end{tabular}
\end{flushleft}
\end{table}

Fortunately, the possibility exists to check the temporal behaviour of the \sws\ data 
with independent \isocam\ observations (see Table\,\ref{tabcam}).
We performed aperture photometry on \cam\ images, which were obtained in two
broad-band filters centred on \about\,7 (LW\,2) and 15\,\um\ (LW\,3), respectively. These 
observations were performed when DEOS was both in the \sws-high (April 14, 1996) 
and in the \sws-low state (September 22, 1997). Table\,\ref{tabcam} summarises the 
photometry results for DEOS and a reference star (EC\,53), expressed in Jy/beam, 
where the `per beam' refers to the aperture size of the \sws. Evidently, the \cam\ data 
also reveal significant variability for DEOS, by a factor of about 1.3 in both filters. 
At 7\,\um, the \sws\ spectra are too noisy to be meaningfully measured, but the \sws\ 
data in the 15\,\um\ region are in excellent agreement with the \cam-value.

\begin{figure}
  \resizebox{\hsize}{!}{\rotatebox{90}{\includegraphics{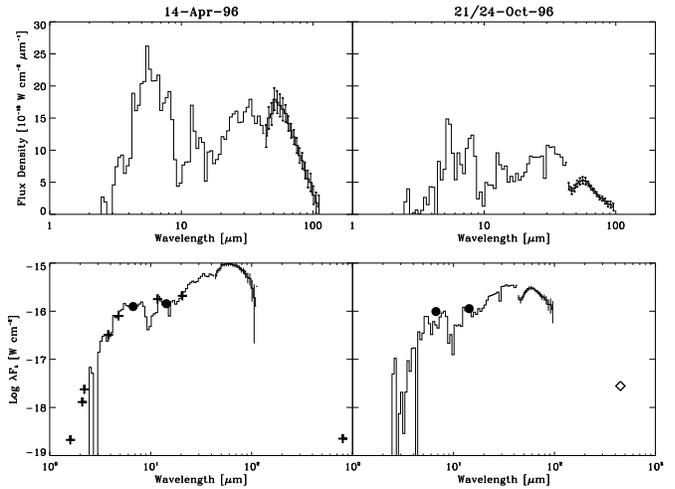}}}
  \caption{SEDs of DEOS in high (left) and low state (right), referring to the dates shown above the
  frames. The \sws\ parts of the data have been directly observed,
  whereas the \lws\ branches result from the spectro-spatial deconvolution modelling. Plus-signs refer
  to data by Hodapp et al. (1996), and are thus not simultaneous with the \iso\ data, and filled circles 
  to the \isocam\ data presented in Table\,\ref{tabcam}; other symbols are as in Fig.\,\ref{smm1modbbfit}.
  The SEDs contain considerable spectral structure. Using the calibration by Whittet (1998)
  the optical depth of the silicate feature at 10\,\um, $\tau_{9.7} \sim 1.3$, would implicate some 
  20 magnitudes of visual extinction to the continuum source}
  \label{deos}
\end{figure}

\begin{figure}
  \resizebox{\hsize}{!}{\rotatebox{90}{\includegraphics{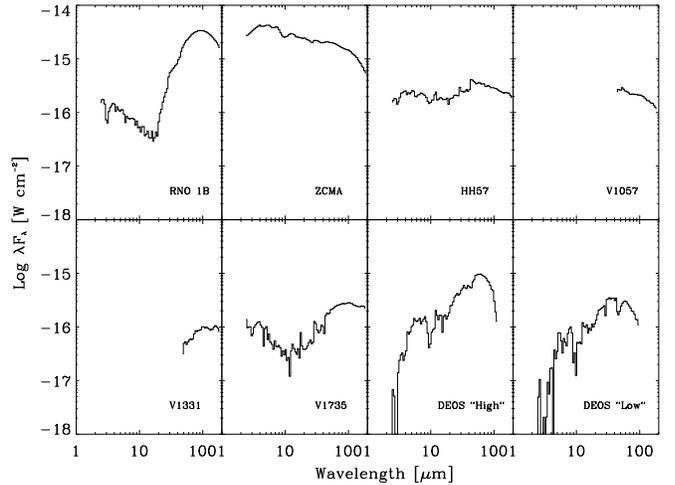}}}
  \caption{\iso\ observations of FUORs resulted in the infrared SEDs of RNO\,1B, Z\,CMa, HH\,57, 
  V\,1057\,Cyg, V\,1331\,Cyg and V\,1735\,Cyg. For reference the SEDs of DEOS are shown to scale}
  \label{fuor2}
\end{figure}

In conclusion, the evidence presented firmly establishes that DEOS had changed its flux in the 
near- to mid-infrared between April and October, 1996. It seems very likely therefore
that also the \lws\ spectra show traces of this variability. In this context we recall that
in the submillimetre SMM\,9/S\,68N is the strongest source. Hence, it seems very likely
that, at the time of the \lws\ observations in October 1996, DEOS was not dominant at the longest 
wavelengths, but might have contributed significantly at the shorter \lws\ wavelengths. 
A longwave fit for S\,68N to the pointed \lws\ observation was properly scaled and subtracted 
and the spectro-spatial deconvolution modelling was done for the residual. This led finally
to the solutions for the SEDs of DEOS presented in Fig.\,\ref{deos}. 

The joining of the {\it observed} \sws\ and {\it modelled} \lws\ spectra for both the
high and the low state is quite spectacular. This indicates that the algorithm 
we have used is indeed capable of providing satisfactory results. We suggest therefore 
that the spectra presented in Fig.\,\ref{deos} represent the most likely SEDs for the 
outburst source DEOS. As far as we are aware, these would be seen for the first time and 
should be helpful in constraining outburst models and, perhaps, in differentiating source 
type (e.g. FUOR vs non-FUOR or EXOR). The far-infrared SEDs of FUORs appear very different
though (Fig.\,\ref{fuor2}).

\subsection{Summarising discussion}  

Tabs.\,\ref{tab_smm_bb_par} and \ref{tab_smm_phys_param} summarise the results of the \lws\ observations and our
spectro-spatial deconvolution modelling. For cool sources, the good wavelength
coverage and resolution of the \lws\ is generally better adapted to estimate the temperature than
the poor resolution and short range of \iras\ data. The source emission is generally optically thick
in the \lws\ spectral range. Compared to the estimates based on \iras\ data
by Hurt \& Barsony (1996) our temperatures of the SMM sources are generally higher, indicative of
a larger number of photons heating the dust. This is also effectively expressed by our larger 
luminosities, although the derived sizes of the \fir\ photospheres (e.g., $\tau_{100 \mu m} \sim 1$)
are generally smaller than the dimensions obtained by Hurt \& Barsony (1996).

We have deduced the observed $\beta$ values from the slopes of the Rayleigh-Jeans part of the source SEDs 
and obtain generally quite shallow distributions, viz. $\mid\!\beta\!\mid \leq 1$. This is in line
with the findings by, e.g., Casali et al. (1993), Testi \& Sargent (1998) and Davis et al. (1999).
However, for a given source, the scatter among different observations is large and it appears doubtful 
that $\beta$ could be a useful diagnostic for SMM-type objects. In fact, the observed flatness 
probably indicates that the assumption of optically thin emission is not valid. This hypothesis
is supported by detailed radiative transfer models, in which the temperature distribution    
is self-consistently calculated. Models were calculated for five sources, for which sufficient 
data were available, viz. SMM\,1, 2, 3, 4 and 9. These models are optically thick out to millimetre 
wavelengths and envelope masses are in the range 2--6\,\msun. 

The luminosities of these SMMs are in the range 10--70\,\lsun. Much or most of this luminosity is 
presumably produced by mass accretion processes. These depend on the centrally accumulated mass. 
It is at present difficult to tell whether the SMMs will end up on the main sequence as low-mass 
or as intermediate mass stars (\lapprox\,1--3\,\msun; the distance too comes in here as a crucial parameter). 
As OH and/or \water\ masers normally are found towards young sources of higher mass, the detection 
of such emission towards SMM\,1 by Rodr\'{\i}guez et al. (1989) and Curiel et al. (1993), respectively, 
would speak in favour of the intermediate mass option, at least for that object. In any case, 
the existence of these massive envelopes around the SMMs argues for the idea that these sources 
are in a very early stage of their development. One can also speculate that most of these
massive envelopes will have become dispersed on a relatively short timescale (\about\,\powten{5}\,yr).

\section{Conclusions}

The main conclusions of this work can be summarised as follows:

\begin{itemize}
\item[$\bullet$] In order to disentangle the contribution of individual sources to the \fir\ emission in
a 8\amin\,$\times$\,8\amin\ map obtained with the \iso-\lws\ a {\it Maximum Likelihood
Method} is introduced. This spectro-spatial deconvolution technique enabled us to restore the \fir\
spectra (50 to 200\,\um) of previously known submillimetre sources (SMMs).
\item[$\bullet$] Observations of SMMs with the \lws\ are advantagous as the data sample the peak of the
spectral energy distributions (SEDs). This permits an accurate determination of the dust temperatures.
These temperatures are generally found to be higher than those obtained from high resolution \iras\ data.
\item[$\bullet$] Fits to the observed SEDs by modified blackbodies reveal that the SMMs are generally 
optically thick at \lws\ wavelengths. 
\item[$\bullet$] In addition, for reasonable assumptions about the grain opacities, the Rayleigh-Jeans part 
of the observed SEDs is significantly flatter than what would be expected for optically thin dust emission.
We interpret this to indicate the SMMs to be optically thick out too much longer wavelengths 
than previously assumed. 
\item[$\bullet$] Self-consistent radiative transfer calculations confirm the correctness of this assertion.
These models of the SMMs are optically thick out to mm-wavelengths. These dust envelopes are massive, 
several \msun, suggesting that these sources are still in their very infancy. 
\item[$\bullet$] The outburst source DEOS has been observed with various instruments aboard \iso\ and at
different times. The source had certainly varied in infrared brightness and we present the full infrared 
spectra, at both high and low states, for the first time.
\end{itemize}

\acknowledgements{We wish to thank the referee, N.J. Evans II, for valuable suggestions which led to 
improvements of the manuscript. The authors from the Stockholm Observatory acknowledge the support by
the Swedish National Space Board.}

\appendix

\section{Spectro-spatial deconvolution: the algorithm}

The aim of the presented algorithm is to restore the SEDs of sources in a crowded field.
To be able to handle spatially undersampled and noisy spectroscopic data we
developed a program based on the {\it Maximum Likelihood Method} (e.g., Lucy 1974).
The present discussion is based on our \lws\ observations, but the method per se should 
of course be more widely applicable.

The first of two basic assumptions is that all sources are point-like, 
which is probably reasonable considering the large beam size of the \lws\ 
(\hpbw\,\about\,70\asec). Secondly, the exact positions of all sources
are assumed known and kept fixed. We then pick an arbitary grid point within the field of the 
\lws-map and try to estimate, at that location and for a fixed frequency, 
the radiative flux which originated from all sources in the field. 
This is done at all locations and for all \lws\ wavelengths in an iterative scheme as
described in the following. 

The photon distribution can be assumed to follow Poisson statistics and, hence, the 
probability to observe $O_{m}$ detector counts at the position $m$ can be written as

\begin{equation}
P_{m} = \frac{e^{-E_{m}}\,E_{m}^{O_{m}}}{O_{m}!}
\end{equation}

where

\begin{equation}
E_{m} = \sum_{s=1}^{N_{S}} \Omega_{m,s}\,S_{s}
\end{equation}

is the expected number of counts at position $m$, arriving from $N_{S}$
sources, each contributing $S_{s}$ counts and being at positions $s$.
$\Omega_{m,s}$ is the diffraction pattern of the instrument (the `beam',
see Fig.\,\ref{lws_beam_size_fit}). Designating the likelihood to observe 
the counts from a map containing $N_{M}$ points by

\begin{equation}
L = P_{1}\,P_{2}\,P_{3}\,\cdots\,P_{N_{M}}
\end{equation}

one can also write

\begin{equation}
\ln L = \sum_{m=1}^{N_{M}} O_{m}\ln E_{m} - E_{m} - \ln O_{m}!
\end{equation}

Solving for the maximum likelihood, viz. 

\begin{equation}
\frac{\partial \ln L}{\partial S_{s}} = 0
\end{equation}

for all positions $s$, or

\begin{equation}
\sum_{m=1}^{N_{M}}\Omega_{m,s}\,\frac{O_{m}}{E_{m}} =
\sum_{m=1}^{N_{M}}\Omega_{m,s}
\end{equation}

From the last expression, a correction factor can be obtained

\begin{equation}
C_{s} = \frac{\sum_{m=1}^{N_{M}} \Omega_{m,s} 
\frac{F_{m}^{O}}{F_{m}^{M}}}
{\sum_{m=1}^{N_{M}} \Omega_{m,s}}
\end{equation}

where $F_{m}^{O}$ is now the observed flux (for a linear detector) at position $m$ and where

\begin{equation}
F_{m}^{M} = \sum_{s=1}^{N_{S}} \Omega_{m,s}\,F_{s}^{S}
\end{equation}

is the expected flux at this position, originating from $N_{S}$ sources 
at positions $s$ and contributing the flux $F_{s}^{S}$. Finally, the iteration is then

\begin{equation}
F_{s}^{S}\left(New\right) = C_{s} F_{s}^{S}\left(Old\right)
\end{equation}

for all wavelengths $\lambda$, yielding the source spectra.

\section{Spectro-spatial deconvolution: test}

\begin{figure}
  \resizebox{\hsize}{!}{\rotatebox{90}{\includegraphics{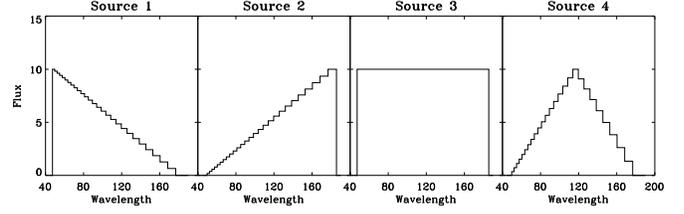}}}
  \caption{The \fir\ spectra of the four point sources included in the test example
  have simple forms in the wavelength range of the \lws\ observations. The 
  binning corresponds, as before, to $R_{\lambda}=20$}
  \label{orgsed}
\end{figure}
\begin{figure}
  \resizebox{\hsize}{!}{\rotatebox{90}{\includegraphics{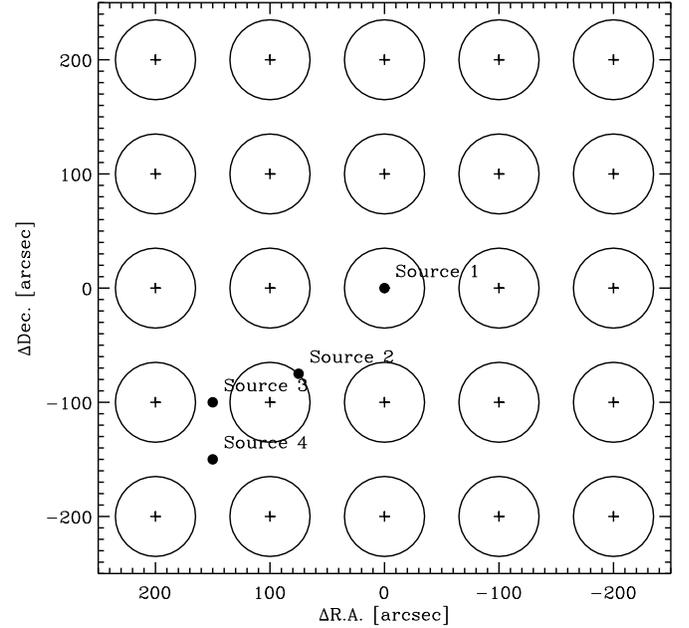}}}
  \caption{Map showing the grid points of the simulated observation, the \hpbw\ of the \lws\ 
  and the positions of the point sources\,1 to 4 having the spectra shown in Fig.\,\ref{orgsed}}
  \label{mapsource}
\end{figure}
\begin{figure}
  \resizebox{\hsize}{!}{\rotatebox{90}{\includegraphics{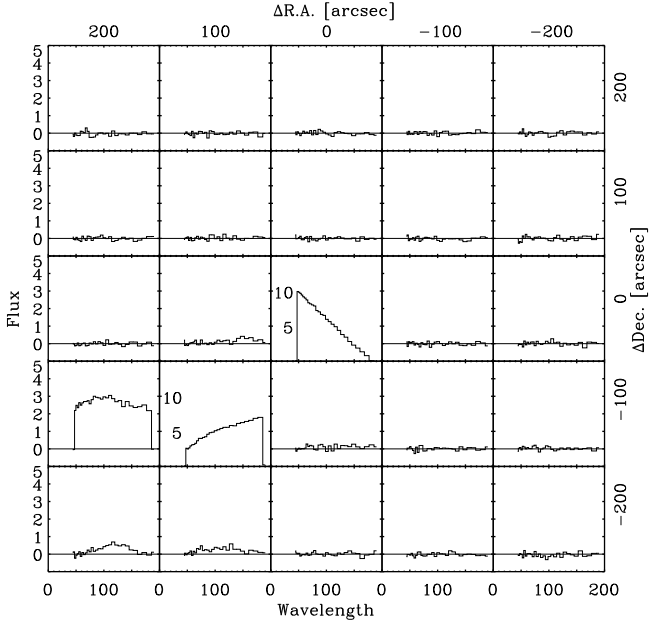}}}
  \caption{The simulated observation of sources\,1 to 4 (with added noise) results 
  in the shown undersampled \lws\ map of spectra (cf. Fig.\,\ref{rawdata}). 
  Notice in particular the very faint and noisy appearance of source\,4}
  \label{mapdata}
\end{figure}
\begin{figure}
  \resizebox{\hsize}{!}{\rotatebox{90}{\includegraphics{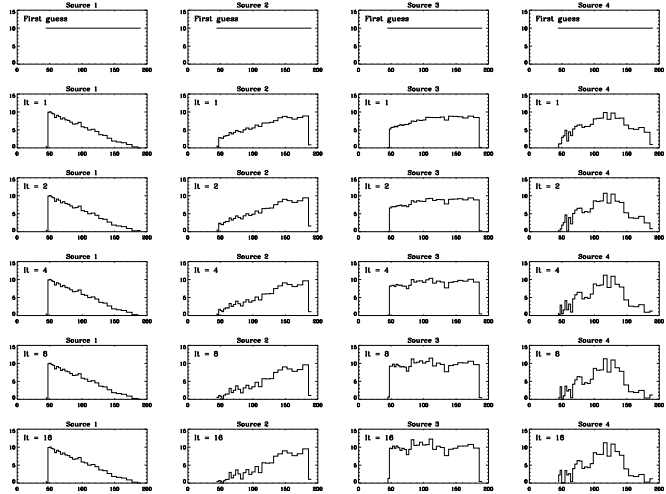}}}
  \caption{The iterative spectro-spatial deconvolution process is shown vertically for 
  the sources\,1 to 4, from left to right. The first row shows the start conditions 
  for the source spectra, viz. simply a constant spectral distribution. Convergence 
  is achieved already after four iterations, shown in the fourth row. Iterating further
  does not improve on the results: see rows 5 and 6, showing the results after 8 and
  16 iterations, respectively}
  \label{itres}
\end{figure}

As an example of a test case, an observation of the \lws\ map was simulated. We
chose four \fir\ spectra, the shape of which is described by simple geometric
figures (positive ramp, negative ramp, rectangular box and triangle: see 
Fig.\,\ref{orgsed}). The point-like sources of these spectra were distributed on 
the \lws\ grid in the manner depicted in Fig.\,\ref{mapsource}, i.e. source\,1
centred on a grid point (which therefore should not be affected by the convolution,
but still by the noise), 
source\,2 on the half-power contour of the \lws\ beam, source\,3 situated half-way 
between two grid points and source\,4 farthest away from any grid point. 

At each wavelength (spectral resolution $R_{\lambda}=20$)
for every map position the sources were convolved with the \lws\ beam of 
Fig.\,\ref{lws_beam_size_fit} and (white) noise was added. This results in the
simulated observations shown in Fig.\,\ref{mapdata}. These ``observed'' maps
(one per spectral bin) were then run thrugh the {\it Maximum Likelihood} 
algorithm for a number of iterations. 

The result of the deconvolution is shown
in Fig.\,\ref{itres}, where the succession of iterations is ordered vertically 
for each of the four sources. The first row shows the start conditions, where we 
``guessed'' a simple constant spectrum in all cases. These straight line spectra
have changed to resemble the true source spectra already after the first run. In this 
particular test case, the deconvolutions have converged after about 4 iterations 
(fourth row), i.e. successive iterations do not give further improvements. 
Apparently, the spectra of particularly source\,1, as expected, but also 
of sources\,2 and 3 are very well reproduced. 
The most difficult test case is source\,4 (cf also Fig.\,\ref{mapdata}), 
which is located farthest away from the map points, and therefore considerably
more affected by the noise. Nevertheless, our method reproduced its spectrum
reasonably well too. 

We conclude that this test procedure was successful, which lends credibility to 
the reliability and potential of the developed spectro-spatial deconvolution algorithm.

\end{document}